\documentclass[aip,reprint,graphicx,twocolumn,superscriptaddress,nofootinbib]{revtex4-2}
\usepackage{amssymb,amsmath,amsbsy,graphicx,xcolor,times,subcaption,marvosym,color,bm,multirow,fontenc}
\usepackage[utf8]{inputenc}
\begin{document}

\title
  {Near-field imaging of plasmonic nanopatch antennas with integrated semiconductor quantum dots}
\author{Vasudevan Iyer}
\affiliation{Center  for  Nanophase  Materials Sciences, Oak Ridge National Laboratory, Oak Ridge, TN 37831, USA}
\author{Yoong Sheng Phang}
\affiliation{University of Georgia, Athens, GA, 30602, USA}
\author{Andrew Butler}
\affiliation{Department of Electrical and Computer Engineering, University of Nebraska-Lincoln, Lincoln, NE 68588, USA}
\author{Jiyang Chen}
\affiliation{Department of Physics and Materials Science, University of Memphis, Memphis, TN, 38152, USA}
\author{Brian Lerner}
\affiliation{Department of Electrical and Computer Engineering, Duke University, Durham, NC 27708, USA}
\author{Christos Argyropoulos}
\affiliation{Department of Electrical and Computer Engineering, University of Nebraska-Lincoln, Lincoln, NE 68588, USA}
\author{Thang Hoang}
\affiliation{Department of Physics and Materials Science, University of Memphis, Memphis, TN, 38152, USA}
\author{Benjamin Lawrie}
\email[]{lawriebj@ornl.gov}
\affiliation{Quantum Science Center, Oak Ridge, TN 37831 USA}
\affiliation{Materials Science and Technology Division, Oak Ridge National Laboratory, Oak Ridge, TN 37831, USA}
\affiliation{Center  for  Nanophase  Materials Sciences, Oak Ridge National Laboratory, Oak Ridge, TN 37831, USA}

\date{\today}

\begin{abstract}
Plasmonic nanopatch antennas that incorporate dielectric gaps hundreds of picometers to several nanometers thick have drawn increasing attention over the past decade because they confine electromagnetic fields to grossly sub-diffraction limited volumes.  Substantial control over the optical properties of excitons and color centers confined within these plasmonic cavities has already been demonstrated with far-field optical spectroscopies, but near-field optical spectroscopies are essential to an improved understanding of the plasmon-emitter interaction at the nanoscale. Here, we characterize the intensity and phase-resolved plasmonic response of isolated nanopatch antennas with cathodoluminescence microscopy. Further, we explore the distinction between optical and electron-beam spectroscopies of coupled plasmon-exciton heterostructures to identify constraints and opportunities for future nanoscale characterization and control of hybrid nanophotonic structures. While we observe substantial Purcell enhancement in time-resolved photoluminescence spectroscopies, negligible Purcell enhancement is observed in cathodoluminescence spectroscopies of hybrid nanophotonic structures.  The substantial differences in measured Purcell enhancement for electron-beam and laser excitation can be understood as a result of the different selection rules for these complementary experiments. These results provide a fundamentally new understanding of near-field plasmon-exciton interactions in nanopatch antennas that is essential to myriad emerging quantum photonic devices.
\end{abstract}


\maketitle 

\section{Introduction}
Plasmonic heterostructures that offer grossly sub-diffraction-limited field confinement have been a topic of intense research for much of the past twenty years\cite{wu2021bright,baumberg2019extreme,maier2003local,maier2006plasmonic,brongersma2000electromagnetic}. This nanoscale field confinement has enabled advances in optical information processing\cite{Zijlstra2009}, photocatalysis\cite{Linic2011}, and plasmon enhanced sensors and spectroscopies\cite{Nie1102,maier2006plasmonic,lee2021quantum}, as well as a variety of studies of plasmonic analogues to cavity quantum electrodynamics\cite{baumberg2019extreme,Benz726,lawrie2012plasmon,fofang2008plexcitonic,Akselrod2014,Akselrod2016,Huang2018}. While conventional lithographically patterned plasmonic nanostructures like split ring resonators\cite{doi:10.1063/1.3194154}, bowtie antennas\cite{Jin2006}, nanodiscs \cite{lawrie2012plasmon}, and other nanopatterned resonators \cite{davidson2016ultrafast} can offer a sufficiently small mode volume to enable the exploration of  weak and strong plasmon-exciton coupling \cite{lawrie2012plasmon,fofang2008plexcitonic,pamu2021broadband}, the emergence of gap plasmon modes that are confined to gaps of order 1 nm has enabled substantially increased flexibility in the control of the optical properties of excitons and color centers confined within the gap\cite{baumberg2019extreme,Benz726,Akselrod2014,Akselrod2016,Huang2018}. Plasmonic nanogap heterostructures have been used to enhance the quantum efficiency of defect and exciton emission in the weak coupling regime\cite{Akselrod2016,Jiang_2014, Hoang2016,Huang2018,Bogdanov2018,Luo2018}, to modify optical properties in the strong coupling regime\cite{kleemann2017strong,chikkaraddy2016single,qin2020revealing}, and for applications ranging from gas sensing\cite{2016Smith}, water splitting\cite{Xu2015}, and carbon capture\cite{Lu2020} to the non-invasive study of nanoparticle dynamics in liquids\cite{Bischak2017}. Electrical tuning of the resonance frequency of nanogap antennas has also been demonstrated as part of the move toward more functional nanophotonic devices\cite{ElecTuning2016}. 

While finite difference time domain (FDTD) simulations and finite element models (FEM) are widely used to simulate the near-field optical response of plasmonic cavities, inhomogeneities in the actual devices ultimately constrain the photonic response. A clear understanding of plasmonic near-field interactions is essential to uncover the structure-function relationship for nanophotonic media. Techniques such as electron energy loss spectroscopy (EELS) and cathodoluminescence (CL) microscopy are well suited to study the near-field optical properties of plasmonic heterostructures\cite{Hachtel2019,Pakeltis:21,hachtel2018polarization} and hybrid plasmon-emitter heterostructures\cite{feldman2018colossal}. In this work, we use CL microscopy to map localized and propagating plasmon modes in the proximity of silver (Ag) nanopatch antennas and analyze the effect of this plasmon mode structure on CdSe quantum dots deposited within the nanopatch antenna (NPA). We compare the results with steady state and time-resolved photoluminescence measurements. Furthermore, we perform FEM simulations that improve our understanding of the plasmonic modes observed in CL. Lastly, we observe interference between electron-beam generated transition radiation and surface plasmon polaritons scattered from the NPA. Notably, while substantial Purcell enhancement is observed in the time-resolved photoluminescence (TRPL) spectra, negligible Purcell enhancement is observed in the CL spectra as a result of different selection rules for electron-beam and optical excitations. Combining CL microscopy of nanoplasmonic modes with PL microscopy of plasmon-exciton interactions is thus essential to understanding plasmon-exciton interactions in nanopatch antennas that are essential to a plethora of emerging applications, such as light absorbers, photodetectors, and photon sources.

\begin{figure}[h]
    \centering
    \includegraphics[width=\columnwidth]{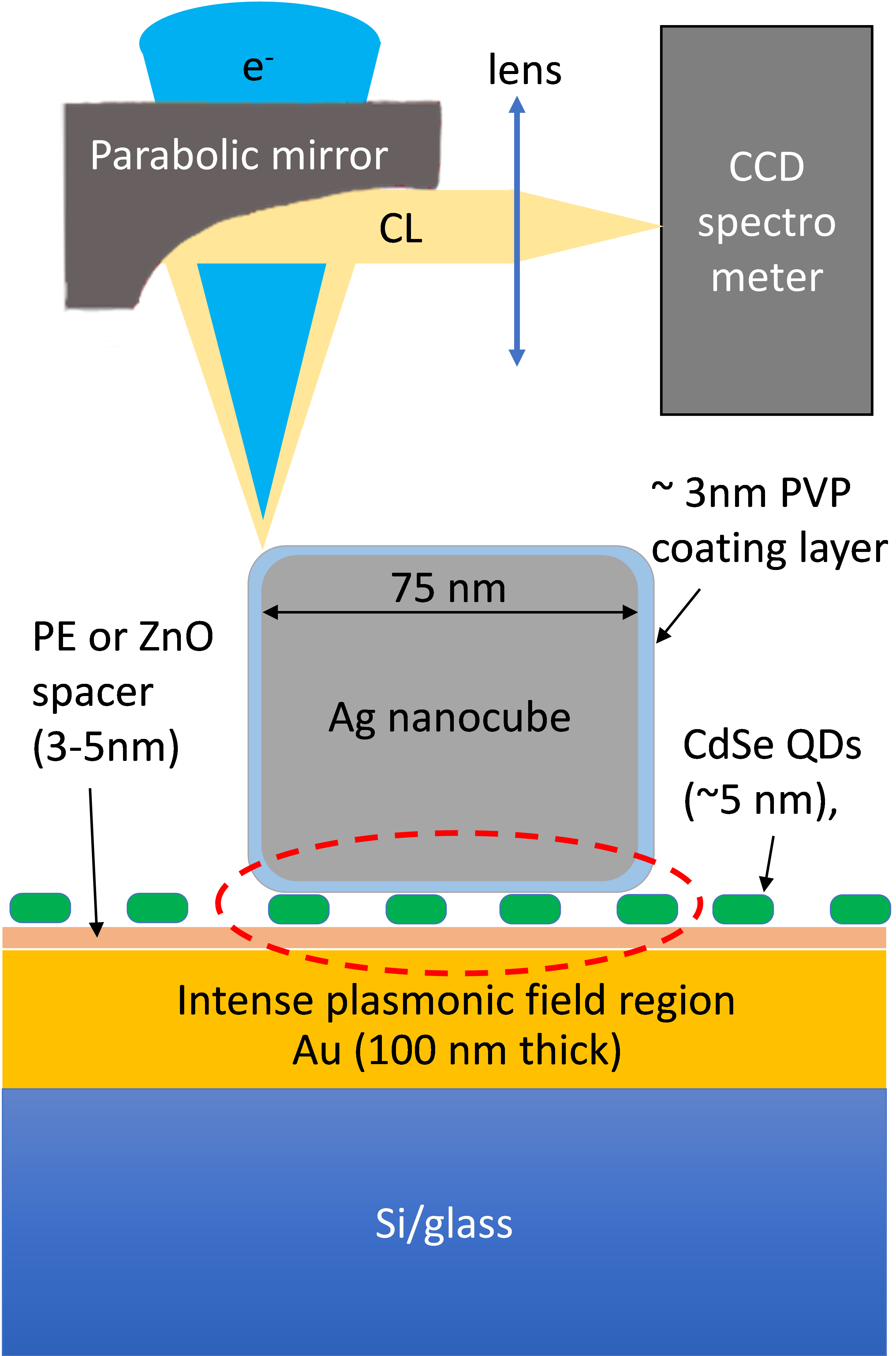}
    \caption{Schematic of nanopatch-antenna heterostructure. QDs are sandwiched between the spacer layer and nanocube. The cathodoluminescence is obtained by exciting the structure with an electron beam in an SEM and using a high NA parabolic mirror to collect light onto a spectrometer. Image is not to scale.}
    \label{fig:sampleSchematic}
\end{figure}

\section{Spectrum Imaging of Individual Nanopatch Antennas}

\begin{figure}[h]
    \centering
    \includegraphics[width=\columnwidth]{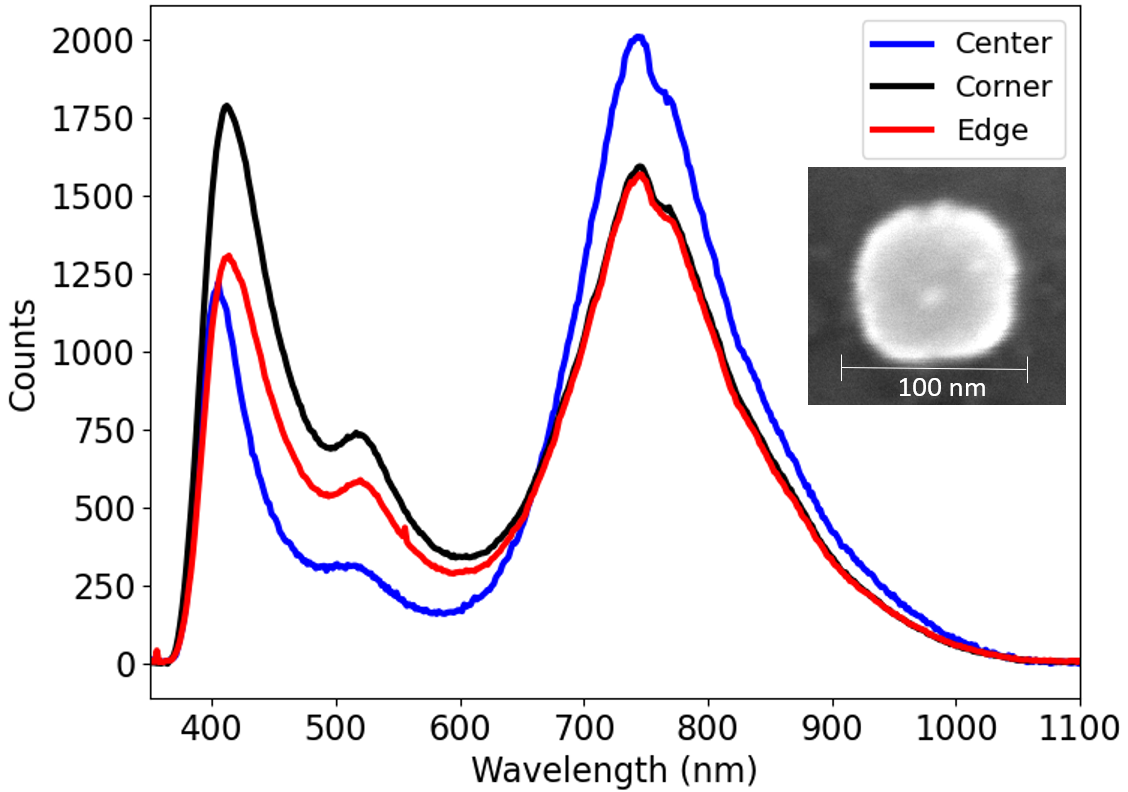}
    \caption{Nanopatch antenna CL spectra for heterostructures with polyelectrolyte spacer layer. Inset shows the SEM image of the cube. The spectra show three peaks corresponding to corner, edge, and gap plasmon modes. The three spectra were obtained with the electron beam positioned at the center, corner and edge, respectively}
    \label{fig:cubeSpectrum}
\end{figure}

\begin{figure*}[!htb]
    \centering
    \hspace{-0.35cm}
    \includegraphics[width=1.01\textwidth]{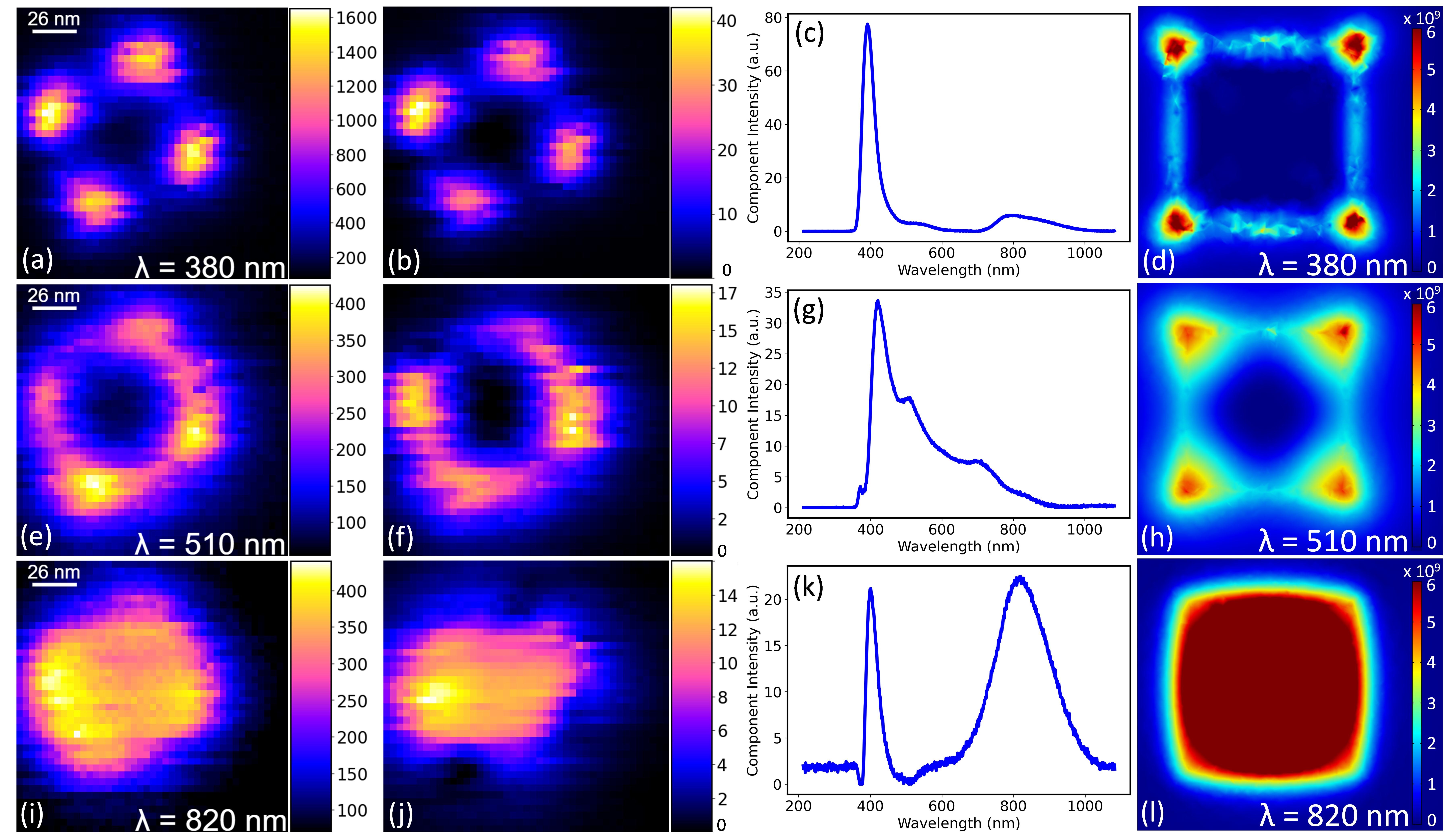}
    \caption{Spectrum images of plasmon resonances supported by Ag nanocube heterostructures with ZnO spacer layer on Au thin films. left column (a,e,i): raw measured CL intensity maps at wavelengths of 380 nm, 510 nm, and 820 nm (with 5 nm bandwidth for each image), second column (b,f,j): The largest three of four NMF decomposition components, third column (c,g,k): associated NMF spectra for each component, right column (d,h,l): FEM simulations of the electric field induced by the electron-beam in the gap between the nanocube and the Au thin film at the same wavelengths used in the left column.}
    \label{fig:modes_rep}
\end{figure*}

The nanopatch heterostructures explored here comprise Ag nanocubes with 75-nm edges dropcast onto a 100-nm gold (Au) thin film.  A 3-5 nm spacer layer is deposited between the Au film and the nanocube to minimize charge transfer and to maximize field confinement within the spacer layer.  For initial spectroscopic mapping of the plasmon modes, a 3 nm ZnO spacer layer was deposited on the Au films on silicon substrates by atomic layer deposition.  For subsequent spectroscopies of coupled plasmon-exciton heterostructures, a peel-and-stick technique was used to transfer gold films with sub-nm surface roughness onto a glass substrate, onto which a polyelectrolyte (PE) spacer layer (3 nm in thickness) was deposited with a dip-coating technique, and CdSe QDs were then spincast before Ag nanocube deposition. A cross-sectional schematic of the fabricated NPA heterostructures and a schematic representation of the CL collection optics are shown in Figure \ref{fig:sampleSchematic}.

The nanocube's plasmonic modes were probed directly by CL microscopy in a FEI Quattro environmental scanning electron microscope (ESEM) with a Delmic Sparc CL collection system. All data described here were acquired with a beam energy of 30 keV, and beam currents of 630-5600 pA were used depending on sample robustness. While the ESEM can operate in a high vacuum or water vapor environment, all data reported here were taken in a background pressure of 40 Pa of water vapor in order to minimize charging and carbon deposition during long spectrum image acquisitions. The NPAs display three prominent plasmon modes centered at wavelengths of $\sim$ 400 nm, 520 nm and 780 nm, as shown in Figure \ref{fig:cubeSpectrum} for a typical nanocube. The inset shows the SEM image of the cube. The mode around $\sim$ 780 nm is strongest when the electron beam is in the center whereas the mode around $\sim$ 400 nm is strongest near the corner. The edge mode is weakly excited at the center. However, the point spectra by themselves provide only a limited understanding of the heterostructure's plasmonic response. The edge and corner modes were previously observed in the near field with EELS\cite{Goris2014} for nanocube assemblies (different than the current nanopatch antenna system), but EELS cannot distinguish the radiative and non radiative modes, a critical requirement for hybrid plasmon-exciton heterostructures.

The spectrum images shown in Figure \ref{fig:modes_rep} depict the spatial distribution of distinct modes supported by the nanocube. The images shown in the left column (Figures \ref{fig:modes_rep}(a), (e), (i)) represent the raw CL intensity at wavelengths of 380, 510, and 820 nm that best represent the corner, edge, and gap plasmon modes. However, simply plotting CL images associated with a limited number of wavelengths tends to obscure the information within the larger spectrum image. Hence, we used non-negative matrix factorization (NMF) to decompose the spectrum image into relevant spectral components. NMF has become a widely used tool for analysis of hyperspectral microscopies because it offers a computationally inexpensive approach to extract sparse, physically relevant data from multidimensional data\cite{konevcna2021revealing,liu2019spatiotemporal}.

\begin{figure*}[!htb]
    \centering
    \hspace{-0.45cm}
    \includegraphics[width=1.01\textwidth]{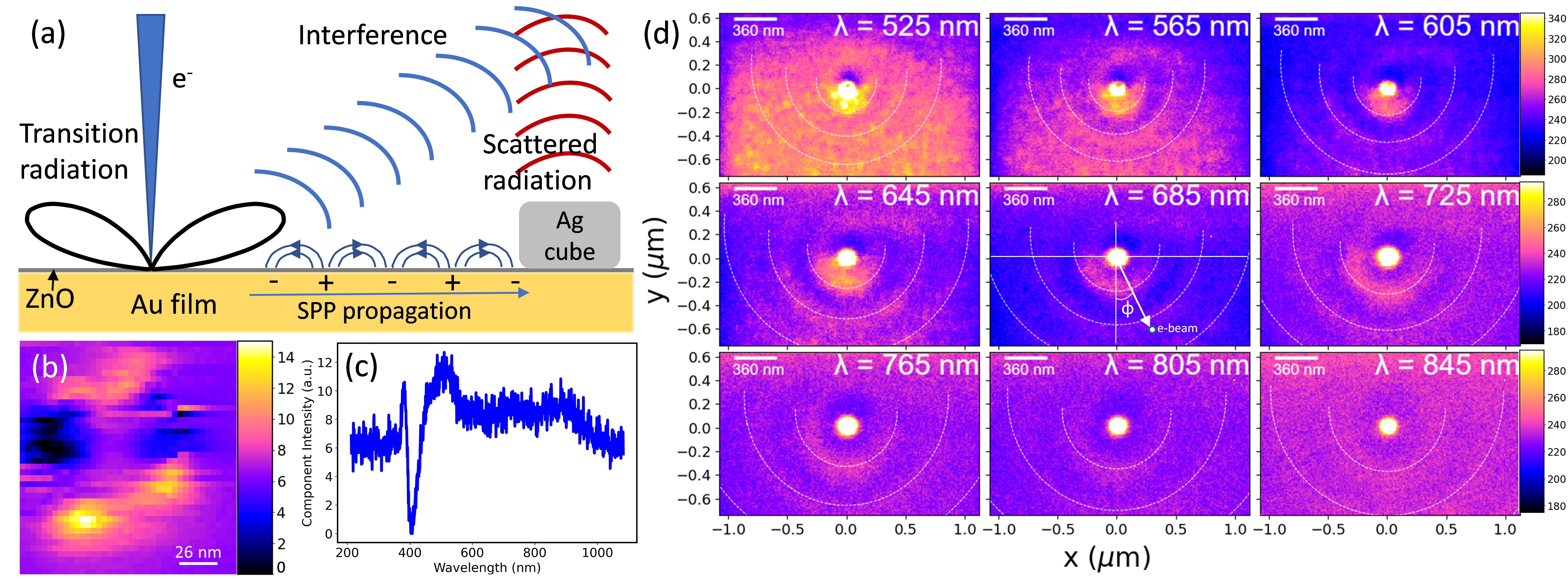}
    \caption{Interference of electron-beam generated transition radiation and scattered surface plasmons from the NPA. (a) Schematic of transition radiation and SPP generation at point of electron-beam incidence, followed by SPP propagation and scattering by NPA. The interference is observed in the far field. (b,c) Final NMF component images from the NMF decomposition shown in Fig. 3. (d) Interference patterns as the electron beam is scanned over a large area surrounding the nanocube. The colorbar is over-saturated near the NPA to enable a clearer image of the weak interference fringes.}
    \label{fig:interference}
\end{figure*}

Figures \ref{fig:modes_rep} (b), (f) and (j) show the three most prominent modes obtained by four-component NMF, which closely match the modes identified in the raw CL intensity maps (the fourth component is described in Fig.\ref{fig:interference}). The NMF spectra for each of the three components are shown in Figures \ref{fig:modes_rep} (c), (g) and (k). Notably, while the NMF decomposition identifies a high quality-factor corner plasmon resonance, the edge mode exhibits some additional spectral complexity, and the gap plasmon mode exhibits a clear bimodal response that is not evident in the direct intensity plot.  These raw intensity plots and NMF decompositions are largely consistent with FEM simulations of the plasmonic field confinement induced by a line-current source emulating an 'electron-beam' (as shown in the right column of Fig. \ref{fig:modes_rep}), though they certainly exhibit the hallmarks of experimental heterogeneities.


\section{Plasmon interference with transition radiation}

Nanoparticles on metallic planes can act as an antenna to couple free space photons with surface plasmon polaritons (SPPs) \cite{Sannomiya2020}. It is crucial to understand this coupling for the current nanopatch antenna system to engineer nanophotonic devices that rely on the out coupling of SPPs to the far field. Notably, the final component of the NMF decomposition shown in Figure~\ref{fig:modes_rep} (shown in Figures \ref{fig:interference} (b) and (c)), is weak compared with the localized modes, and it is very broadband except for an artifact associated with the strong corner modes near 400 nm. It is consistent with broadband SPP modes scattering at the nanocube, but this high resolution image provides no detailed understanding of the SPP propagation on the Au film. Thus, we acquired spectrum images with larger fields of view to characterize interference between transition radiation and SPP CL.  This interference allows for near-field imaging of the phase of surface plasmon polaritons before they scatter off the nanocube\cite{kuttge2009local,Hachtel2019}.  A clear understanding of SPP propagation combined with the above imaging of nanogap plasmon modes is critical to understanding the composite plasmon-exciton interaction. This understanding is also essential to the ultimate development of dissipative driven entanglement schemes that rely on coupling multiple emitters to shared plasmonic reservoirs\cite{dumitrescu2017antibunching,li2019resonance}. Broadband transition radiation is generated when the image-charge-induced dipole in the metal is destroyed as the electron enters the metal. The electron beam also launches SPPs in the metal, which propagate outward. The SPP is scattered into free space photons when it interacts with the nanopatch antenna. A schematic of this process is shown in Figure \ref{fig:interference} (a). This scattered radiation has a characteristic phase relative to the transition radiation depending on the distance of the cube from the electron beam. Therefore, the scattered beam and transition radiation can interfere to produce the interferograms shown in Figure \ref{fig:interference} (d). 

The radial distance which satisfies the interference condition is given by \cite{Sannomiya2020}:
\begin{equation}
R = \frac{\Phi_{osc}+2m\pi}{k_{spp}+k\ sin\ \theta \ cos\ \phi}+\Delta,
\end{equation}
where $\Phi_{osc}$ is the phase of the field scattered from the NPA, \textit{m} is an integer, $k_{spp}$ is the plasmon wave vector that is equal to $k\sqrt{\frac{\epsilon_1\epsilon_2}{\epsilon_1+\epsilon_2}}$, where $\epsilon_1$ and $\epsilon_2$ are the permittivities of ZnO and metal film, respectively, $\phi$ is the azimuthal angle between the NPA
and point of electron-beam incidence and $\Delta$ is the phase arising due to the distance between the plasmon origin and the NPA. It was previously reported that the phase $\Phi_{osc}$ is zero away from plasmon resonances \cite{Sannomiya2020}, and hence due to the high energy of the corner and edge modes ($<$ 520 nm), we set $\Phi_{osc} = 0$ in the wavelength range of 525 to 825 nm, where we observed interference. The parabolic mirror collects all angles $\theta$ from 0 to $\sim$ 90 $^{\circ}$ and we display the fitting result for $\theta = 45^{\circ}$ as white dashed lines in Figure \ref{fig:interference} (d). The thickness of the measured fringes is understood by changing $\theta$ from 0 to $\sim$ 90$^{\circ}$. We plot the calculated fringes for $-110^{\circ} < \phi < 110^{\circ}$, due to the reduced experimental signal to noise at higher angles. $\Delta$ was chosen as $0.17\lambda_{spp}$, $0.35\lambda_{spp}$ and $0.7\lambda_{spp}$ for the first, second and third interference fringes for all images. Good agreement is obtained between the experimental and theoretical results. This technique can be used for any plasmonic nanoparticle shape to obtain the phase delay and scattering phase.

\begin{figure}[h]
    \centering
    \includegraphics[width=\columnwidth]{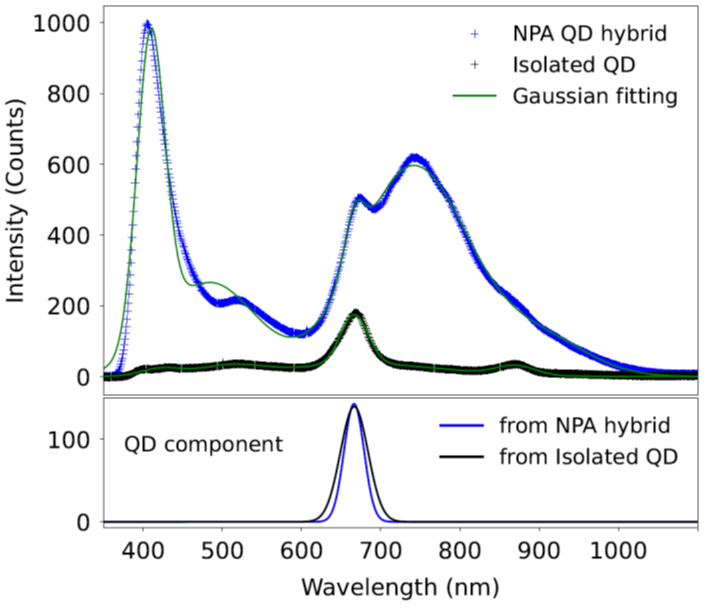}
    \caption{CL emission of CdSe QDs embedded in the NPA hybrid and on PE layer. The spectra are fit with five Gaussian peaks and the QD emission contribution is extracted and shown in the bottom panel for excitation on the NPA and isolated QD, respectively.}
    \label{fig:polymerEnhancement}
\end{figure}

\begin{figure}[h]
    \captionsetup[subfigure]{labelformat=empty}
    \centering
    \begin{subfigure}{1\columnwidth}
        \centering
        \includegraphics[width=\columnwidth]{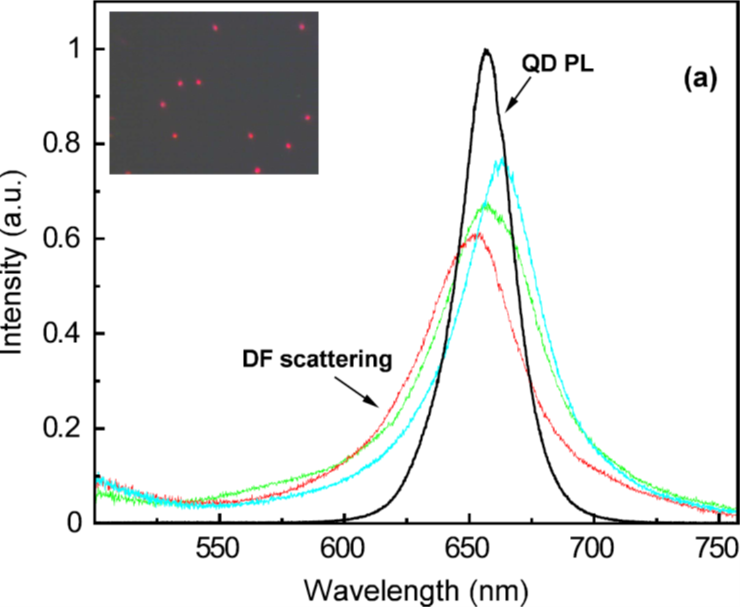}
    \end{subfigure}
    
    \begin{subfigure}{1\columnwidth}
        \centering
        \includegraphics[width=\columnwidth]{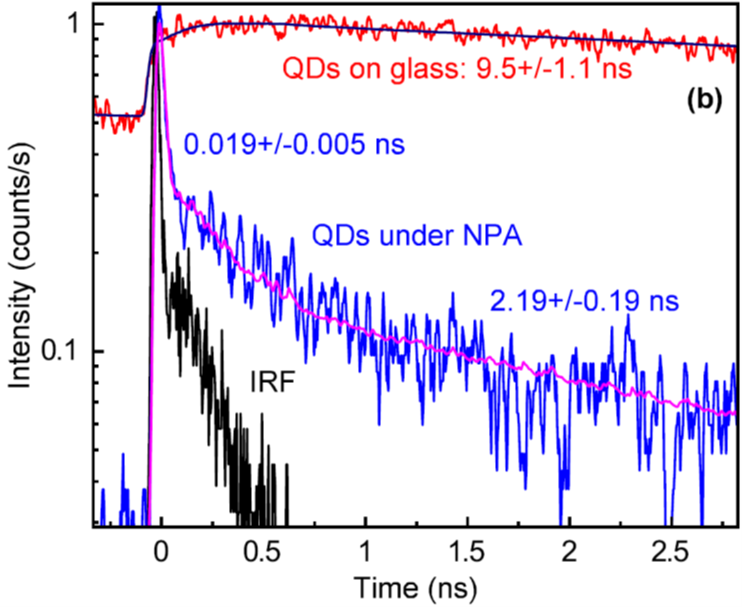}
    \end{subfigure}
    \caption{(a) Quantum dot PL emission and dark-field scattering spectra of nanopatch antennas. The inset displays a dark-field camera image of the NPAs under white light illumination. Individual NPAs were well separated and easily distinguished for optical spectroscopies. (b) Time-resolved PL of quantum dot emission. The decreasing decay time implies a Purcell enhancement effect.}
    \label{fig:PL-TRPL}
\end{figure}

\section{Coupled nanopatch quantum dot heterostructures}

While CL microscopy enables clear spectrum imaging of nanogap plasmon modes in these NPAs by using ZnO as spacer layer, the PE layer was moderately deformed by the electron beam during image acquisition.  This deformation did not substantially change the optical properties of the nanogap plasmon modes as evidenced by the spectra in Figure \ref{fig:cubeSpectrum}, but it did modify the morphology of the substrate sufficiently to make drift compensation problematic for long acquisition-time spectrum images.  As a result, spectrum images presented in Figure \ref{fig:modes_rep} are for the isolated nanocubes on Au film coated silicon substrates with ZnO spacer layers. Below, we compare the CL emission for two different cases where  QDs are deposited on a PE spacer layer (on a Au film) with or without nanocubes. We find that in contrast to the PE surface, it is a challenge to adhere the CdSe QDs directly on a ZnO surface (prior to the nanocube deposition). It is also noted that, to date, there has been no careful comparison of the PL and CL of hybrid structures designed to exhibit large Purcell effects.

Point spectra for the CdSe quantum dots dropcast on the PE spacer layer and for the hybrid nanopatch-CdSe quantum dot heterostructure are shown in Figure \ref{fig:polymerEnhancement}. The data are fit with five Gaussian peaks. The quantum dot contribution to the plasmon-exciton heterostructure is extracted as one of those fit components and plotted separately in the lower panel of Figure \ref{fig:polymerEnhancement}. Compared with previous reports of quantum dot photoluminescence enhancement by nanogap plasmon modes\cite{Akselrod2016}, the cathodoluminescence enhancement compared with the CL of isolated QDs is negligible. This is consistent with a relatively limited literature to date exploring Purcell enhancement of exciton and defect emission by plasmonic heterostructures, where only minimal Purcell enhancement was observed. \cite{feldman2018colossal,jiang2020tailoring,yanagimoto2020purcell}. This is not necessarily surprising, because the electron beam-pumped system exhibits a very different interaction cross-section and different selection rules than the laser-pumped system.

In order to help understand the selection rules associated with electron- and laser-driven plasmon-exciton heterostructures, we acquired complementary optical dark-field (DF) scattering spectra and TRPL spectra in an optical microscope shown in Figures \ref{fig:PL-TRPL}(a) and (b) respectively. The DF scattering spectra of several individual NPAs shown in Fig. \ref{fig:PL-TRPL}(a) exhibit a small variation in the peak wavelength that is consistent with small variation in the nanocube size or gap thickness. Nevertheless, the NPAs' peak resonances show an excellent spectral overlap with the QD emission at around 650 nm.  Notably, the measured DF scattering under optical excitation exhibits much simpler mode structure than the measured cathodoluminescence because the near-field electron-beam excitation is capable of exciting optically forbidden transitions. Figure \ref{fig:PL-TRPL}(b) compares the TRPL of QDs on a glass slide and in an NPA. Similar to the previously observed Purcell enhancement effect in the NPA platform, the decay time of QDs in the NPA is significantly reduced as compared to the intrinsic decay of the same QDs on a glass slide\cite{Akselrod2014,Akselrod2016,Hoang2015UF,Bogdanov2018}. The decay of QDs on a glass slide is well fitted with a single exponential component with a lifetime of 9.5 ns. When the QDs are coupled to the NPA, the decay curve is fitted with a biexponential function deconvolved with the instrument response, resulting in a fast decay component of 0.019 ns and a slower component of 2.19 ns. The fast component is attributed to the decay time of the QDs that are well coupled to the NPA and the slower component is originated from QDs on the Au film, away from the nanocube but within the laser excitation spot. The reduction in the decay of the QDs on a Au film is a result of the nonradiative quenching effect. As it has been shown previously that the decrease in lifetime of the QDs in the NPA is accompanied by a simultaneous increase in the PL intensity \cite{Akselrod2014,Akselrod2016,Hoang2015UF,Bogdanov2018}, the ratio $\tau_{glass}/\tau_{NPA}\sim$ 500 represents the Purcell enhancement factor.

\section{Conclusion}
The TRPL measurements suggest that the heterostructure is well designed with measured Purcell enhancement consistent with the literature, and the samples were transferred in vacuum from the TRPL measurement to the CL measurement, so minimal environmental degradation is expected between measurements. Therefore, the negligible measured Purcell enhancement with electron beam excitation indicates a more fundamental difference between the two measurement modalities. The near-field electron-beam excitation of the NPA excites additional modes than are excited optically, resulting in a broader red-shifted composite NPA response, as shown in Figure \ref{fig:polymerEnhancement}. Though the QD emission lies within the broadband envelope, the electron beam excitation of modes that are far from resonant with the QD result in substantially reduced plasmon-exciton interactions in the composite system. 

However,  while it seems clear that CL characterization   of   hybrid   plasmon-exciton   heterostructures  may  not  yield  the  same  Purcell enhancement  measured  in  PL  spectroscopies, CL  microscopy  is an indispensable tool for mapping  localized and  propagating  plasmonic  modes  of  technologically relevant nanopatch antenna structures with  few  nanometer  spatial  resolution. Further, techniques such as electron-beam induced deposition can be used to pattern plasmonic heterostructures in the SEM\cite{winkler2017direct,iyer2021situ} near individual defects or excitons after imaging with CL microscopy, and exert in situ control over plasmon-exciton interactions subject to the understanding that the composite system response will be different under laser excitation than electron beam excitation.  Indeed, pump-probe spectroscopies that combine laser and electron beam excitation of the sample should allow for completely integrated experiments that probe the near-field properties of the hybrid heterostructure with an electron beam excitation while probing the resonantly excited state with a laser excitation. Ultimately, this level of integrated manipulation and characterization of hybrid nanophotonic systems  is critical to the development of various recently proposed dissipative-driven entanglement schemes\cite{dumitrescu2017antibunching,li2019resonance}.

\section{Methods}

\textbf{CL microscopy.} 
CL microscopy was performed in a FEI Quattro environmental scanning electron microscope with a Delmic Sparc cathodoluminescence collection system. A beam energy of 30 keV, and beam currents of 630-5600 pA were used, depending  on  sample  robustness. The sample chamber was maintained at a pressure of 40 Pa water vapor. CL spectra were collected using a parabolic mirror with high numerical aperture of 0.9 and sent to a spectrometer (Andor Kymera 193i) equipped with a 150 line/mm grating and an Andor Newton CCD camera. Spectra were acquired using dwell times of 0.2-2 s per pixel, and spectrum images were acquired by scanning the electron beam while acquiring a CL spectrum at each pixel. Reference images were taken every 2-20 s as necessary during the spectrum image acquisition to allow for drift correction.\\

\textbf{Optical characterization.} Optical responses of NPA samples were characterized by dark-field scattering and PL measurements through a customized Nikon LV 150N microscope with a 3D nanometer-resolution translation stage (Newport, model 9063). The modified microscope has a halogen lamp for white light illumination and a modified optical path for laser excitation (Coherent laser at 475-nm, 80MHz, 150fs) and PL collection through a bright/dark field 100X microscope objective lens. Dark-field scattering or PL signal from individual NPAs was filtered by a pinhole at a focal imaging plane before entering the spectrometer (Horiba iHR550) and CCD camera (Horiba Jobin-Yvon Synapse). A set of appropriate short-pass (for laser excitation) and long-pass (for PL detection) spectral filters were used for the QD excitation and PL collection. For the decay time measurements of the QDs, a time-correlated single photon counting setup was used. The PL emission from the QDs was collected by the objective lens and sent through the spectrometer for either spectral analysis or through a side exit to be guided into a fast-timing avalanche photodiode for temporal analysis. The signal collected by the photodiode was analyzed by a single photon counting module (PicoHarp 300). Final lifetimes were obtained from fits to the data deconvolved with the instrument response function by using the EasyTau software (PicoQuant).\\

\textbf{FEM simulation.} Three-dimensional (3D) FEM simulations of the silver (Ag) plasmonic nanopatch antenna structures were performed using Comsol Multiphysics®. The dimensions of the nanocube were identical to the experiments and a spacer layer with thickness of 5 nm was used. The model was surrounded with a spherical scattering boundary condition that prevented backscattering of the electromagnetic waves from the simulation boundaries. A line current was placed vertically 10 nm above the center of the nanocube to simulate the electron beam.\cite{Sutter2021, Sutter2021Opto} The induced field plots shown in Figure 3 were obtained by running the simulation twice, once with the line current in vacuum and once with the line current in the presence of the nanopatch antenna, each time using the same mesh. The induced fields were obtained by using the formula: $E_{ind}=E_{ant}-E_{vac}$, where $E_{ant}$ is the field calculated with the nanopatch antenna structure and Evac is the field calculated in vacuum. All employed materials were simulated using realistic refractive index data.\cite{Stelling2017,PhysRevB.6.4370}\\


\section{acknowledgement}

This research was carried out at the Center for Nanophase Materials Sciences (CNMS), which is sponsored at ORNL by the Scientific User Facilities Division, Office of Basic Energy Sciences, U.S. Department of Energy. Support at ORNL for hybrid quantum photonic systems was provided by the U.S. Department of Energy, Office of Science, National Quantum Information Science Research Centers, Quantum Science Center. CA acknowledges support by Office of Naval Research Young Investigator Program (ONR-YIP) under grant no. N00014-19-1-2384. TH acknowledges support from the National Science Foundation (NSF) (Grant no. DMR-1709612).


\begin{thebibliography}{51}%
\makeatletter
\providecommand \@ifxundefined [1]{%
 \@ifx{#1\undefined}
}%
\providecommand \@ifnum [1]{%
 \ifnum #1\expandafter \@firstoftwo
 \else \expandafter \@secondoftwo
 \fi
}%
\providecommand \@ifx [1]{%
 \ifx #1\expandafter \@firstoftwo
 \else \expandafter \@secondoftwo
 \fi
}%
\providecommand \natexlab [1]{#1}%
\providecommand \enquote  [1]{``#1''}%
\providecommand \bibnamefont  [1]{#1}%
\providecommand \bibfnamefont [1]{#1}%
\providecommand \citenamefont [1]{#1}%
\providecommand \href@noop [0]{\@secondoftwo}%
\providecommand \href [0]{\begingroup \@sanitize@url \@href}%
\providecommand \@href[1]{\@@startlink{#1}\@@href}%
\providecommand \@@href[1]{\endgroup#1\@@endlink}%
\providecommand \@sanitize@url [0]{\catcode `\\12\catcode `\$12\catcode
  `\&12\catcode `\#12\catcode `\^12\catcode `\_12\catcode `\%12\relax}%
\providecommand \@@startlink[1]{}%
\providecommand \@@endlink[0]{}%
\providecommand \url  [0]{\begingroup\@sanitize@url \@url }%
\providecommand \@url [1]{\endgroup\@href {#1}{\urlprefix }}%
\providecommand \urlprefix  [0]{URL }%
\providecommand \Eprint [0]{\href }%
\providecommand \doibase [0]{https://doi.org/}%
\providecommand \selectlanguage [0]{\@gobble}%
\providecommand \bibinfo  [0]{\@secondoftwo}%
\providecommand \bibfield  [0]{\@secondoftwo}%
\providecommand \translation [1]{[#1]}%
\providecommand \BibitemOpen [0]{}%
\providecommand \bibitemStop [0]{}%
\providecommand \bibitemNoStop [0]{.\EOS\space}%
\providecommand \EOS [0]{\spacefactor3000\relax}%
\providecommand \BibitemShut  [1]{\csname bibitem#1\endcsname}%
\let\auto@bib@innerbib\@empty
\bibitem [{\citenamefont {Wu}, \citenamefont {Yan},\ and\ \citenamefont
  {Lalanne}(2021)}]{wu2021bright}%
  \BibitemOpen
  \bibfield  {author} {\bibinfo {author} {\bibfnamefont {T.}~\bibnamefont
  {Wu}}, \bibinfo {author} {\bibfnamefont {W.}~\bibnamefont {Yan}},\ and\
  \bibinfo {author} {\bibfnamefont {P.}~\bibnamefont {Lalanne}},\ }\bibfield
  {title} {\enquote {\bibinfo {title} {Bright plasmons with cubic nanometer
  mode volumes through mode hybridization},}\ }\href@noop {} {\bibfield
  {journal} {\bibinfo  {journal} {ACS Photonics}\ }\textbf {\bibinfo {volume}
  {8}},\ \bibinfo {pages} {307--314} (\bibinfo {year} {2021})}\BibitemShut
  {NoStop}%
\bibitem [{\citenamefont {Baumberg}\ \emph {et~al.}(2019)\citenamefont
  {Baumberg}, \citenamefont {Aizpurua}, \citenamefont {Mikkelsen},\ and\
  \citenamefont {Smith}}]{baumberg2019extreme}%
  \BibitemOpen
  \bibfield  {author} {\bibinfo {author} {\bibfnamefont {J.~J.}\ \bibnamefont
  {Baumberg}}, \bibinfo {author} {\bibfnamefont {J.}~\bibnamefont {Aizpurua}},
  \bibinfo {author} {\bibfnamefont {M.~H.}\ \bibnamefont {Mikkelsen}},\ and\
  \bibinfo {author} {\bibfnamefont {D.~R.}\ \bibnamefont {Smith}},\ }\bibfield
  {title} {\enquote {\bibinfo {title} {Extreme nanophotonics from ultrathin
  metallic gaps},}\ }\href@noop {} {\bibfield  {journal} {\bibinfo  {journal}
  {Nature materials}\ }\textbf {\bibinfo {volume} {18}},\ \bibinfo {pages}
  {668--678} (\bibinfo {year} {2019})}\BibitemShut {NoStop}%
\bibitem [{\citenamefont {Maier}\ \emph {et~al.}(2003)\citenamefont {Maier},
  \citenamefont {Kik}, \citenamefont {Atwater}, \citenamefont {Meltzer},
  \citenamefont {Harel}, \citenamefont {Koel},\ and\ \citenamefont
  {Requicha}}]{maier2003local}%
  \BibitemOpen
  \bibfield  {author} {\bibinfo {author} {\bibfnamefont {S.~A.}\ \bibnamefont
  {Maier}}, \bibinfo {author} {\bibfnamefont {P.~G.}\ \bibnamefont {Kik}},
  \bibinfo {author} {\bibfnamefont {H.~A.}\ \bibnamefont {Atwater}}, \bibinfo
  {author} {\bibfnamefont {S.}~\bibnamefont {Meltzer}}, \bibinfo {author}
  {\bibfnamefont {E.}~\bibnamefont {Harel}}, \bibinfo {author} {\bibfnamefont
  {B.~E.}\ \bibnamefont {Koel}},\ and\ \bibinfo {author} {\bibfnamefont
  {A.~A.}\ \bibnamefont {Requicha}},\ }\bibfield  {title} {\enquote {\bibinfo
  {title} {Local detection of electromagnetic energy transport below the
  diffraction limit in metal nanoparticle plasmon waveguides},}\ }\href@noop {}
  {\bibfield  {journal} {\bibinfo  {journal} {Nature materials}\ }\textbf
  {\bibinfo {volume} {2}},\ \bibinfo {pages} {229--232} (\bibinfo {year}
  {2003})}\BibitemShut {NoStop}%
\bibitem [{\citenamefont {Maier}(2006)}]{maier2006plasmonic}%
  \BibitemOpen
  \bibfield  {author} {\bibinfo {author} {\bibfnamefont {S.~A.}\ \bibnamefont
  {Maier}},\ }\bibfield  {title} {\enquote {\bibinfo {title} {Plasmonic field
  enhancement and sers in the effective mode volume picture},}\ }\href@noop {}
  {\bibfield  {journal} {\bibinfo  {journal} {Optics Express}\ }\textbf
  {\bibinfo {volume} {14}},\ \bibinfo {pages} {1957--1964} (\bibinfo {year}
  {2006})}\BibitemShut {NoStop}%
\bibitem [{\citenamefont {Brongersma}, \citenamefont {Hartman},\ and\
  \citenamefont {Atwater}(2000)}]{brongersma2000electromagnetic}%
  \BibitemOpen
  \bibfield  {author} {\bibinfo {author} {\bibfnamefont {M.~L.}\ \bibnamefont
  {Brongersma}}, \bibinfo {author} {\bibfnamefont {J.~W.}\ \bibnamefont
  {Hartman}},\ and\ \bibinfo {author} {\bibfnamefont {H.~A.}\ \bibnamefont
  {Atwater}},\ }\bibfield  {title} {\enquote {\bibinfo {title} {Electromagnetic
  energy transfer and switching in nanoparticle chain arrays below the
  diffraction limit},}\ }\href@noop {} {\bibfield  {journal} {\bibinfo
  {journal} {Physical Review B}\ }\textbf {\bibinfo {volume} {62}},\ \bibinfo
  {pages} {R16356} (\bibinfo {year} {2000})}\BibitemShut {NoStop}%
\bibitem [{\citenamefont {Zijlstra}, \citenamefont {Chon},\ and\ \citenamefont
  {Gu}(2009)}]{Zijlstra2009}%
  \BibitemOpen
  \bibfield  {author} {\bibinfo {author} {\bibfnamefont {P.}~\bibnamefont
  {Zijlstra}}, \bibinfo {author} {\bibfnamefont {J.~W.~M.}\ \bibnamefont
  {Chon}},\ and\ \bibinfo {author} {\bibfnamefont {M.}~\bibnamefont {Gu}},\
  }\bibfield  {title} {\enquote {\bibinfo {title} {Five-dimensional optical
  recording mediated by surface plasmons in gold nanorods},}\ }\href
  {https://doi.org/10.1038/nature08053} {\bibfield  {journal} {\bibinfo
  {journal} {Nature}\ }\textbf {\bibinfo {volume} {459}},\ \bibinfo {pages}
  {410--413} (\bibinfo {year} {2009})}\BibitemShut {NoStop}%
\bibitem [{\citenamefont {Linic}, \citenamefont {Christopher},\ and\
  \citenamefont {Ingram}(2011)}]{Linic2011}%
  \BibitemOpen
  \bibfield  {author} {\bibinfo {author} {\bibfnamefont {S.}~\bibnamefont
  {Linic}}, \bibinfo {author} {\bibfnamefont {P.}~\bibnamefont {Christopher}},\
  and\ \bibinfo {author} {\bibfnamefont {D.~B.}\ \bibnamefont {Ingram}},\
  }\bibfield  {title} {\enquote {\bibinfo {title} {Plasmonic-metal
  nanostructures for efficient conversion of solar to chemical energy},}\
  }\href {https://doi.org/10.1038/nmat3151} {\bibfield  {journal} {\bibinfo
  {journal} {Nature Materials}\ }\textbf {\bibinfo {volume} {10}},\ \bibinfo
  {pages} {911--921} (\bibinfo {year} {2011})}\BibitemShut {NoStop}%
\bibitem [{\citenamefont {Nie}\ and\ \citenamefont {Emory}(1997)}]{Nie1102}%
  \BibitemOpen
  \bibfield  {author} {\bibinfo {author} {\bibfnamefont {S.}~\bibnamefont
  {Nie}}\ and\ \bibinfo {author} {\bibfnamefont {S.~R.}\ \bibnamefont
  {Emory}},\ }\bibfield  {title} {\enquote {\bibinfo {title} {Probing single
  molecules and single nanoparticles by surface-enhanced raman scattering},}\
  }\href {https://doi.org/10.1126/science.275.5303.1102} {\bibfield  {journal}
  {\bibinfo  {journal} {Science}\ }\textbf {\bibinfo {volume} {275}},\ \bibinfo
  {pages} {1102--1106} (\bibinfo {year} {1997})},\ \Eprint
  {https://arxiv.org/abs/https://science.sciencemag.org/content/275/5303/1102.full.pdf}
  {https://science.sciencemag.org/content/275/5303/1102.full.pdf} \BibitemShut
  {NoStop}%
\bibitem [{\citenamefont {Lee}\ \emph {et~al.}(2021)\citenamefont {Lee},
  \citenamefont {Lawrie}, \citenamefont {Pooser}, \citenamefont {Lee},
  \citenamefont {Rockstuhl},\ and\ \citenamefont {Tame}}]{lee2021quantum}%
  \BibitemOpen
  \bibfield  {author} {\bibinfo {author} {\bibfnamefont {C.}~\bibnamefont
  {Lee}}, \bibinfo {author} {\bibfnamefont {B.}~\bibnamefont {Lawrie}},
  \bibinfo {author} {\bibfnamefont {R.}~\bibnamefont {Pooser}}, \bibinfo
  {author} {\bibfnamefont {K.-G.}\ \bibnamefont {Lee}}, \bibinfo {author}
  {\bibfnamefont {C.}~\bibnamefont {Rockstuhl}},\ and\ \bibinfo {author}
  {\bibfnamefont {M.}~\bibnamefont {Tame}},\ }\bibfield  {title} {\enquote
  {\bibinfo {title} {Quantum plasmonic sensors},}\ }\href
  {https://doi.org/10.1021/acs.chemrev.0c01028} {\bibfield  {journal} {\bibinfo
   {journal} {Chemical Reviews}\ }\textbf {\bibinfo {volume} {121}},\ \bibinfo
  {pages} {4743--4804} (\bibinfo {year} {2021})},\ \bibinfo {note} {pMID:
  33787252}\BibitemShut {NoStop}%
\bibitem [{\citenamefont {Benz}\ \emph {et~al.}(2016)\citenamefont {Benz},
  \citenamefont {Schmidt}, \citenamefont {Dreismann}, \citenamefont
  {Chikkaraddy}, \citenamefont {Zhang}, \citenamefont {Demetriadou},
  \citenamefont {Carnegie}, \citenamefont {Ohadi}, \citenamefont {de~Nijs},
  \citenamefont {Esteban}, \citenamefont {Aizpurua},\ and\ \citenamefont
  {Baumberg}}]{Benz726}%
  \BibitemOpen
  \bibfield  {author} {\bibinfo {author} {\bibfnamefont {F.}~\bibnamefont
  {Benz}}, \bibinfo {author} {\bibfnamefont {M.~K.}\ \bibnamefont {Schmidt}},
  \bibinfo {author} {\bibfnamefont {A.}~\bibnamefont {Dreismann}}, \bibinfo
  {author} {\bibfnamefont {R.}~\bibnamefont {Chikkaraddy}}, \bibinfo {author}
  {\bibfnamefont {Y.}~\bibnamefont {Zhang}}, \bibinfo {author} {\bibfnamefont
  {A.}~\bibnamefont {Demetriadou}}, \bibinfo {author} {\bibfnamefont
  {C.}~\bibnamefont {Carnegie}}, \bibinfo {author} {\bibfnamefont
  {H.}~\bibnamefont {Ohadi}}, \bibinfo {author} {\bibfnamefont
  {B.}~\bibnamefont {de~Nijs}}, \bibinfo {author} {\bibfnamefont
  {R.}~\bibnamefont {Esteban}}, \bibinfo {author} {\bibfnamefont
  {J.}~\bibnamefont {Aizpurua}},\ and\ \bibinfo {author} {\bibfnamefont
  {J.~J.}\ \bibnamefont {Baumberg}},\ }\bibfield  {title} {\enquote {\bibinfo
  {title} {Single-molecule optomechanics in
  {\textquotedblleft}picocavities{\textquotedblright}},}\ }\href
  {https://doi.org/10.1126/science.aah5243} {\bibfield  {journal} {\bibinfo
  {journal} {Science}\ }\textbf {\bibinfo {volume} {354}},\ \bibinfo {pages}
  {726--729} (\bibinfo {year} {2016})},\ \Eprint
  {https://arxiv.org/abs/https://science.sciencemag.org/content/354/6313/726.full.pdf}
  {https://science.sciencemag.org/content/354/6313/726.full.pdf} \BibitemShut
  {NoStop}%
\bibitem [{\citenamefont {Lawrie}\ \emph {et~al.}(2012)\citenamefont {Lawrie},
  \citenamefont {Kim}, \citenamefont {Norton},\ and\ \citenamefont
  {Haglund~Jr}}]{lawrie2012plasmon}%
  \BibitemOpen
  \bibfield  {author} {\bibinfo {author} {\bibfnamefont {B.~J.}\ \bibnamefont
  {Lawrie}}, \bibinfo {author} {\bibfnamefont {K.-W.}\ \bibnamefont {Kim}},
  \bibinfo {author} {\bibfnamefont {D.~P.}\ \bibnamefont {Norton}},\ and\
  \bibinfo {author} {\bibfnamefont {R.~F.}\ \bibnamefont {Haglund~Jr}},\
  }\bibfield  {title} {\enquote {\bibinfo {title} {Plasmon--exciton
  hybridization in zno quantum-well al nanodisc heterostructures},}\
  }\href@noop {} {\bibfield  {journal} {\bibinfo  {journal} {Nano letters}\
  }\textbf {\bibinfo {volume} {12}},\ \bibinfo {pages} {6152--6157} (\bibinfo
  {year} {2012})}\BibitemShut {NoStop}%
\bibitem [{\citenamefont {Fofang}\ \emph {et~al.}(2008)\citenamefont {Fofang},
  \citenamefont {Park}, \citenamefont {Neumann}, \citenamefont {Mirin},
  \citenamefont {Nordlander},\ and\ \citenamefont
  {Halas}}]{fofang2008plexcitonic}%
  \BibitemOpen
  \bibfield  {author} {\bibinfo {author} {\bibfnamefont {N.~T.}\ \bibnamefont
  {Fofang}}, \bibinfo {author} {\bibfnamefont {T.-H.}\ \bibnamefont {Park}},
  \bibinfo {author} {\bibfnamefont {O.}~\bibnamefont {Neumann}}, \bibinfo
  {author} {\bibfnamefont {N.~A.}\ \bibnamefont {Mirin}}, \bibinfo {author}
  {\bibfnamefont {P.}~\bibnamefont {Nordlander}},\ and\ \bibinfo {author}
  {\bibfnamefont {N.~J.}\ \bibnamefont {Halas}},\ }\bibfield  {title} {\enquote
  {\bibinfo {title} {Plexcitonic nanoparticles: plasmon- exciton coupling in
  nanoshell- j-aggregate complexes},}\ }\href@noop {} {\bibfield  {journal}
  {\bibinfo  {journal} {Nano letters}\ }\textbf {\bibinfo {volume} {8}},\
  \bibinfo {pages} {3481--3487} (\bibinfo {year} {2008})}\BibitemShut {NoStop}%
\bibitem [{\citenamefont {Akselrod}\ \emph {et~al.}(2014)\citenamefont
  {Akselrod}, \citenamefont {Hoang}, \citenamefont {Cirac{\`i}}, \citenamefont
  {Fang}, \citenamefont {Huang}, \citenamefont {Smith},\ and\ \citenamefont
  {Mikkelsen}}]{Akselrod2014}%
  \BibitemOpen
  \bibfield  {author} {\bibinfo {author} {\bibfnamefont {C.}~\bibnamefont
  {Akselrod}, \bibfnamefont {Gleb M.and~Argyropoulos}}, \bibinfo {author}
  {\bibfnamefont {T.~B.}\ \bibnamefont {Hoang}}, \bibinfo {author}
  {\bibfnamefont {C.}~\bibnamefont {Cirac{\`i}}}, \bibinfo {author}
  {\bibfnamefont {C.}~\bibnamefont {Fang}}, \bibinfo {author} {\bibfnamefont
  {J.}~\bibnamefont {Huang}}, \bibinfo {author} {\bibfnamefont {D.~R.}\
  \bibnamefont {Smith}},\ and\ \bibinfo {author} {\bibfnamefont {M.~H.}\
  \bibnamefont {Mikkelsen}},\ }\bibfield  {title} {\enquote {\bibinfo {title}
  {Probing the mechanisms of large purcell enhancement in plasmonic
  nanoantennas},}\ }\href {https://doi.org/10.1038/nphoton.2014.228} {\bibfield
   {journal} {\bibinfo  {journal} {Nature Photonics}\ }\textbf {\bibinfo
  {volume} {8}},\ \bibinfo {pages} {835--840} (\bibinfo {year}
  {2014})}\BibitemShut {NoStop}%
\bibitem [{\citenamefont {Akselrod}\ \emph {et~al.}(2016)\citenamefont
  {Akselrod}, \citenamefont {Weidman}, \citenamefont {Li}, \citenamefont
  {Argyropoulos}, \citenamefont {Tisdale},\ and\ \citenamefont
  {Mikkelsen}}]{Akselrod2016}%
  \BibitemOpen
  \bibfield  {author} {\bibinfo {author} {\bibfnamefont {G.~M.}\ \bibnamefont
  {Akselrod}}, \bibinfo {author} {\bibfnamefont {M.~C.}\ \bibnamefont
  {Weidman}}, \bibinfo {author} {\bibfnamefont {Y.}~\bibnamefont {Li}},
  \bibinfo {author} {\bibfnamefont {C.}~\bibnamefont {Argyropoulos}}, \bibinfo
  {author} {\bibfnamefont {W.~A.}\ \bibnamefont {Tisdale}},\ and\ \bibinfo
  {author} {\bibfnamefont {M.~H.}\ \bibnamefont {Mikkelsen}},\ }\bibfield
  {title} {\enquote {\bibinfo {title} {Efficient nanosecond photoluminescence
  from infrared pbs quantum dots coupled to plasmonic nanoantennas},}\ }\href
  {https://doi.org/10.1021/acsphotonics.6b00357} {\bibfield  {journal}
  {\bibinfo  {journal} {ACS Photonics}\ }\textbf {\bibinfo {volume} {3}},\
  \bibinfo {pages} {1741--1746} (\bibinfo {year} {2016})}\BibitemShut {NoStop}%
\bibitem [{\citenamefont {Huang}\ \emph {et~al.}(2018)\citenamefont {Huang},
  \citenamefont {Akselrod}, \citenamefont {Ming}, \citenamefont {Kong},\ and\
  \citenamefont {Mikkelsen}}]{Huang2018}%
  \BibitemOpen
  \bibfield  {author} {\bibinfo {author} {\bibfnamefont {J.}~\bibnamefont
  {Huang}}, \bibinfo {author} {\bibfnamefont {G.~M.}\ \bibnamefont {Akselrod}},
  \bibinfo {author} {\bibfnamefont {T.}~\bibnamefont {Ming}}, \bibinfo {author}
  {\bibfnamefont {J.}~\bibnamefont {Kong}},\ and\ \bibinfo {author}
  {\bibfnamefont {M.~H.}\ \bibnamefont {Mikkelsen}},\ }\bibfield  {title}
  {\enquote {\bibinfo {title} {Tailored emission spectrum of 2d semiconductors
  using plasmonic nanocavities},}\ }\href
  {https://doi.org/10.1021/acsphotonics.7b01085} {\bibfield  {journal}
  {\bibinfo  {journal} {ACS Photonics}\ }\textbf {\bibinfo {volume} {5}},\
  \bibinfo {pages} {552--558} (\bibinfo {year} {2018})}\BibitemShut {NoStop}%
\bibitem [{\citenamefont {Cubukcu}\ \emph {et~al.}(2009)\citenamefont
  {Cubukcu}, \citenamefont {Zhang}, \citenamefont {Park}, \citenamefont
  {Bartal},\ and\ \citenamefont {Zhang}}]{doi:10.1063/1.3194154}%
  \BibitemOpen
  \bibfield  {author} {\bibinfo {author} {\bibfnamefont {E.}~\bibnamefont
  {Cubukcu}}, \bibinfo {author} {\bibfnamefont {S.}~\bibnamefont {Zhang}},
  \bibinfo {author} {\bibfnamefont {Y.-S.}\ \bibnamefont {Park}}, \bibinfo
  {author} {\bibfnamefont {G.}~\bibnamefont {Bartal}},\ and\ \bibinfo {author}
  {\bibfnamefont {X.}~\bibnamefont {Zhang}},\ }\bibfield  {title} {\enquote
  {\bibinfo {title} {Split ring resonator sensors for infrared detection of
  single molecular monolayers},}\ }\href {https://doi.org/10.1063/1.3194154}
  {\bibfield  {journal} {\bibinfo  {journal} {Applied Physics Letters}\
  }\textbf {\bibinfo {volume} {95}},\ \bibinfo {pages} {043113} (\bibinfo
  {year} {2009})}\BibitemShut {NoStop}%
\bibitem [{\citenamefont {Jin}\ and\ \citenamefont {Xu}(2006)}]{Jin2006}%
  \BibitemOpen
  \bibfield  {author} {\bibinfo {author} {\bibfnamefont {E.~X.}\ \bibnamefont
  {Jin}}\ and\ \bibinfo {author} {\bibfnamefont {X.}~\bibnamefont {Xu}},\
  }\bibfield  {title} {\enquote {\bibinfo {title} {Plasmonic effects in
  near-field optical transmission enhancement through a single bowtie-shaped
  aperture},}\ }\href {https://doi.org/10.1007/s00340-006-2237-7} {\bibfield
  {journal} {\bibinfo  {journal} {Applied Physics B}\ }\textbf {\bibinfo
  {volume} {84}},\ \bibinfo {pages} {3--9} (\bibinfo {year}
  {2006})}\BibitemShut {NoStop}%
\bibitem [{\citenamefont {Davidson}\ \emph {et~al.}(2016)\citenamefont
  {Davidson}, \citenamefont {Yanchenko}, \citenamefont {Ziegler}, \citenamefont
  {Avanesyan}, \citenamefont {Lawrie},\ and\ \citenamefont
  {Haglund~Jr}}]{davidson2016ultrafast}%
  \BibitemOpen
  \bibfield  {author} {\bibinfo {author} {\bibfnamefont {R.~B.}\ \bibnamefont
  {Davidson}}, \bibinfo {author} {\bibfnamefont {A.}~\bibnamefont {Yanchenko}},
  \bibinfo {author} {\bibfnamefont {J.~I.}\ \bibnamefont {Ziegler}}, \bibinfo
  {author} {\bibfnamefont {S.~M.}\ \bibnamefont {Avanesyan}}, \bibinfo {author}
  {\bibfnamefont {B.~J.}\ \bibnamefont {Lawrie}},\ and\ \bibinfo {author}
  {\bibfnamefont {R.~F.}\ \bibnamefont {Haglund~Jr}},\ }\bibfield  {title}
  {\enquote {\bibinfo {title} {Ultrafast plasmonic control of second harmonic
  generation},}\ }\href@noop {} {\bibfield  {journal} {\bibinfo  {journal} {ACS
  Photonics}\ }\textbf {\bibinfo {volume} {3}},\ \bibinfo {pages} {1477--1481}
  (\bibinfo {year} {2016})}\BibitemShut {NoStop}%
\bibitem [{\citenamefont {Pamu}\ \emph {et~al.}(2021)\citenamefont {Pamu},
  \citenamefont {Lawrie}, \citenamefont {Khomami},\ and\ \citenamefont
  {Mukherjee}}]{pamu2021broadband}%
  \BibitemOpen
  \bibfield  {author} {\bibinfo {author} {\bibfnamefont {R.}~\bibnamefont
  {Pamu}}, \bibinfo {author} {\bibfnamefont {B.~J.}\ \bibnamefont {Lawrie}},
  \bibinfo {author} {\bibfnamefont {B.}~\bibnamefont {Khomami}},\ and\ \bibinfo
  {author} {\bibfnamefont {D.}~\bibnamefont {Mukherjee}},\ }\bibfield  {title}
  {\enquote {\bibinfo {title} {Broadband plasmonic photocurrent enhancement
  from photosystem i assembled with tailored arrays of au and ag nanodisks},}\
  }\href@noop {} {\bibfield  {journal} {\bibinfo  {journal} {ACS Applied Nano
  Materials}\ }\textbf {\bibinfo {volume} {4}},\ \bibinfo {pages} {1209--1219}
  (\bibinfo {year} {2021})}\BibitemShut {NoStop}%
\bibitem [{\citenamefont {Jiang}\ \emph {et~al.}(2014)\citenamefont {Jiang},
  \citenamefont {Shao}, \citenamefont {Yin}, \citenamefont {Liu}, \citenamefont
  {Song}, \citenamefont {Zhu},\ and\ \citenamefont {Xu}}]{Jiang_2014}%
  \BibitemOpen
  \bibfield  {author} {\bibinfo {author} {\bibfnamefont {T.-T.}\ \bibnamefont
  {Jiang}}, \bibinfo {author} {\bibfnamefont {W.-J.}\ \bibnamefont {Shao}},
  \bibinfo {author} {\bibfnamefont {N.-Q.}\ \bibnamefont {Yin}}, \bibinfo
  {author} {\bibfnamefont {L.}~\bibnamefont {Liu}}, \bibinfo {author}
  {\bibfnamefont {J.-L.-Q.}\ \bibnamefont {Song}}, \bibinfo {author}
  {\bibfnamefont {L.-X.}\ \bibnamefont {Zhu}},\ and\ \bibinfo {author}
  {\bibfnamefont {X.-L.}\ \bibnamefont {Xu}},\ }\bibfield  {title} {\enquote
  {\bibinfo {title} {Enhanced photoluminescence of {CdSe} quantum dots by the
  coupling of ag nanocube and ag film},}\ }\href
  {https://doi.org/10.1088/1674-1056/23/8/086102} {\bibfield  {journal}
  {\bibinfo  {journal} {Chinese Physics B}\ }\textbf {\bibinfo {volume} {23}},\
  \bibinfo {pages} {086102} (\bibinfo {year} {2014})}\BibitemShut {NoStop}%
\bibitem [{\citenamefont {Hoang}, \citenamefont {Akselrod},\ and\ \citenamefont
  {Mikkelsen}(2016)}]{Hoang2016}%
  \BibitemOpen
  \bibfield  {author} {\bibinfo {author} {\bibfnamefont {T.~B.}\ \bibnamefont
  {Hoang}}, \bibinfo {author} {\bibfnamefont {G.~M.}\ \bibnamefont
  {Akselrod}},\ and\ \bibinfo {author} {\bibfnamefont {M.~H.}\ \bibnamefont
  {Mikkelsen}},\ }\bibfield  {title} {\enquote {\bibinfo {title} {Ultrafast
  room-temperature single photon emission from quantum dots coupled to
  plasmonic nanocavities},}\ }\href
  {https://doi.org/10.1021/acs.nanolett.5b03724} {\bibfield  {journal}
  {\bibinfo  {journal} {Nano Letters}\ }\textbf {\bibinfo {volume} {16}},\
  \bibinfo {pages} {270--275} (\bibinfo {year} {2016})}\BibitemShut {NoStop}%
\bibitem [{\citenamefont {Bogdanov}\ \emph {et~al.}(2018)\citenamefont
  {Bogdanov}, \citenamefont {Shalaginov}, \citenamefont {Lagutchev},
  \citenamefont {Chiang}, \citenamefont {Shah}, \citenamefont {Baburin},
  \citenamefont {Ryzhikov}, \citenamefont {Rodionov}, \citenamefont
  {Kildishev}, \citenamefont {Boltasseva},\ and\ \citenamefont
  {Shalaev}}]{Bogdanov2018}%
  \BibitemOpen
  \bibfield  {author} {\bibinfo {author} {\bibfnamefont {S.~I.}\ \bibnamefont
  {Bogdanov}}, \bibinfo {author} {\bibfnamefont {M.~Y.}\ \bibnamefont
  {Shalaginov}}, \bibinfo {author} {\bibfnamefont {A.~S.}\ \bibnamefont
  {Lagutchev}}, \bibinfo {author} {\bibfnamefont {C.-C.}\ \bibnamefont
  {Chiang}}, \bibinfo {author} {\bibfnamefont {D.}~\bibnamefont {Shah}},
  \bibinfo {author} {\bibfnamefont {A.~S.}\ \bibnamefont {Baburin}}, \bibinfo
  {author} {\bibfnamefont {I.~A.}\ \bibnamefont {Ryzhikov}}, \bibinfo {author}
  {\bibfnamefont {I.~A.}\ \bibnamefont {Rodionov}}, \bibinfo {author}
  {\bibfnamefont {A.~V.}\ \bibnamefont {Kildishev}}, \bibinfo {author}
  {\bibfnamefont {A.}~\bibnamefont {Boltasseva}},\ and\ \bibinfo {author}
  {\bibfnamefont {V.~M.}\ \bibnamefont {Shalaev}},\ }\bibfield  {title}
  {\enquote {\bibinfo {title} {Ultrabright room-temperature sub-nanosecond
  emission from single nitrogen-vacancy centers coupled to nanopatch
  antennas},}\ }\href {https://doi.org/10.1021/acs.nanolett.8b01415} {\bibfield
   {journal} {\bibinfo  {journal} {Nano Letters}\ }\textbf {\bibinfo {volume}
  {18}},\ \bibinfo {pages} {4837--4844} (\bibinfo {year} {2018})}\BibitemShut
  {NoStop}%
\bibitem [{\citenamefont {Luo}\ \emph {et~al.}(2018)\citenamefont {Luo},
  \citenamefont {Shepard}, \citenamefont {Ardelean}, \citenamefont {Rhodes},
  \citenamefont {Kim}, \citenamefont {Barmak}, \citenamefont {Hone},\ and\
  \citenamefont {Strauf}}]{Luo2018}%
  \BibitemOpen
  \bibfield  {author} {\bibinfo {author} {\bibfnamefont {Y.}~\bibnamefont
  {Luo}}, \bibinfo {author} {\bibfnamefont {G.~D.}\ \bibnamefont {Shepard}},
  \bibinfo {author} {\bibfnamefont {J.~V.}\ \bibnamefont {Ardelean}}, \bibinfo
  {author} {\bibfnamefont {D.~A.}\ \bibnamefont {Rhodes}}, \bibinfo {author}
  {\bibfnamefont {B.}~\bibnamefont {Kim}}, \bibinfo {author} {\bibfnamefont
  {K.}~\bibnamefont {Barmak}}, \bibinfo {author} {\bibfnamefont {J.~C.}\
  \bibnamefont {Hone}},\ and\ \bibinfo {author} {\bibfnamefont
  {S.}~\bibnamefont {Strauf}},\ }\bibfield  {title} {\enquote {\bibinfo {title}
  {Deterministic coupling of site-controlled quantum emitters in monolayer wse2
  to plasmonic nanocavities},}\ }\href
  {https://doi.org/10.1038/s41565-018-0275-z} {\bibfield  {journal} {\bibinfo
  {journal} {Nature Nanotechnology}\ }\textbf {\bibinfo {volume} {13}},\
  \bibinfo {pages} {1137--1142} (\bibinfo {year} {2018})}\BibitemShut {NoStop}%
\bibitem [{\citenamefont {Kleemann}\ \emph {et~al.}(2017)\citenamefont
  {Kleemann}, \citenamefont {Chikkaraddy}, \citenamefont {Alexeev},
  \citenamefont {Kos}, \citenamefont {Carnegie}, \citenamefont {Deacon},
  \citenamefont {de~Pury}, \citenamefont {Gro{\ss}e}, \citenamefont {de~Nijs},
  \citenamefont {Mertens}, \citenamefont {Tartakovskii},\ and\ \citenamefont
  {Baumberg}}]{kleemann2017strong}%
  \BibitemOpen
  \bibfield  {author} {\bibinfo {author} {\bibfnamefont {M.-E.}\ \bibnamefont
  {Kleemann}}, \bibinfo {author} {\bibfnamefont {R.}~\bibnamefont
  {Chikkaraddy}}, \bibinfo {author} {\bibfnamefont {E.~M.}\ \bibnamefont
  {Alexeev}}, \bibinfo {author} {\bibfnamefont {D.}~\bibnamefont {Kos}},
  \bibinfo {author} {\bibfnamefont {C.}~\bibnamefont {Carnegie}}, \bibinfo
  {author} {\bibfnamefont {W.}~\bibnamefont {Deacon}}, \bibinfo {author}
  {\bibfnamefont {A.~C.}\ \bibnamefont {de~Pury}}, \bibinfo {author}
  {\bibfnamefont {C.}~\bibnamefont {Gro{\ss}e}}, \bibinfo {author}
  {\bibfnamefont {B.}~\bibnamefont {de~Nijs}}, \bibinfo {author} {\bibfnamefont
  {J.}~\bibnamefont {Mertens}}, \bibinfo {author} {\bibfnamefont {A.~I.}\
  \bibnamefont {Tartakovskii}},\ and\ \bibinfo {author} {\bibfnamefont {J.~J.}\
  \bibnamefont {Baumberg}},\ }\bibfield  {title} {\enquote {\bibinfo {title}
  {Strong-coupling of wse2 in ultra-compact plasmonic nanocavities at room
  temperature},}\ }\href {https://doi.org/10.1038/s41467-017-01398-3}
  {\bibfield  {journal} {\bibinfo  {journal} {Nature Communications}\ }\textbf
  {\bibinfo {volume} {8}},\ \bibinfo {pages} {1296} (\bibinfo {year}
  {2017})}\BibitemShut {NoStop}%
\bibitem [{\citenamefont {Chikkaraddy}\ \emph {et~al.}(2016)\citenamefont
  {Chikkaraddy}, \citenamefont {De~Nijs}, \citenamefont {Benz}, \citenamefont
  {Barrow}, \citenamefont {Scherman}, \citenamefont {Rosta}, \citenamefont
  {Demetriadou}, \citenamefont {Fox}, \citenamefont {Hess},\ and\ \citenamefont
  {Baumberg}}]{chikkaraddy2016single}%
  \BibitemOpen
  \bibfield  {author} {\bibinfo {author} {\bibfnamefont {R.}~\bibnamefont
  {Chikkaraddy}}, \bibinfo {author} {\bibfnamefont {B.}~\bibnamefont
  {De~Nijs}}, \bibinfo {author} {\bibfnamefont {F.}~\bibnamefont {Benz}},
  \bibinfo {author} {\bibfnamefont {S.~J.}\ \bibnamefont {Barrow}}, \bibinfo
  {author} {\bibfnamefont {O.~A.}\ \bibnamefont {Scherman}}, \bibinfo {author}
  {\bibfnamefont {E.}~\bibnamefont {Rosta}}, \bibinfo {author} {\bibfnamefont
  {A.}~\bibnamefont {Demetriadou}}, \bibinfo {author} {\bibfnamefont
  {P.}~\bibnamefont {Fox}}, \bibinfo {author} {\bibfnamefont {O.}~\bibnamefont
  {Hess}},\ and\ \bibinfo {author} {\bibfnamefont {J.~J.}\ \bibnamefont
  {Baumberg}},\ }\bibfield  {title} {\enquote {\bibinfo {title}
  {Single-molecule strong coupling at room temperature in plasmonic
  nanocavities},}\ }\href@noop {} {\bibfield  {journal} {\bibinfo  {journal}
  {Nature}\ }\textbf {\bibinfo {volume} {535}},\ \bibinfo {pages} {127--130}
  (\bibinfo {year} {2016})}\BibitemShut {NoStop}%
\bibitem [{\citenamefont {Qin}\ \emph {et~al.}(2020)\citenamefont {Qin},
  \citenamefont {Chen}, \citenamefont {Zhang}, \citenamefont {Zhang},
  \citenamefont {Blaikie}, \citenamefont {Ding},\ and\ \citenamefont
  {Qiu}}]{qin2020revealing}%
  \BibitemOpen
  \bibfield  {author} {\bibinfo {author} {\bibfnamefont {J.}~\bibnamefont
  {Qin}}, \bibinfo {author} {\bibfnamefont {Y.-H.}\ \bibnamefont {Chen}},
  \bibinfo {author} {\bibfnamefont {Z.}~\bibnamefont {Zhang}}, \bibinfo
  {author} {\bibfnamefont {Y.}~\bibnamefont {Zhang}}, \bibinfo {author}
  {\bibfnamefont {R.~J.}\ \bibnamefont {Blaikie}}, \bibinfo {author}
  {\bibfnamefont {B.}~\bibnamefont {Ding}},\ and\ \bibinfo {author}
  {\bibfnamefont {M.}~\bibnamefont {Qiu}},\ }\bibfield  {title} {\enquote
  {\bibinfo {title} {Revealing strong plasmon-exciton coupling between nanogap
  resonators and two-dimensional semiconductors at ambient conditions},}\
  }\href@noop {} {\bibfield  {journal} {\bibinfo  {journal} {Physical review
  letters}\ }\textbf {\bibinfo {volume} {124}},\ \bibinfo {pages} {063902}
  (\bibinfo {year} {2020})}\BibitemShut {NoStop}%
\bibitem [{\citenamefont {Powell}\ \emph {et~al.}(2016)\citenamefont {Powell},
  \citenamefont {Coles}, \citenamefont {Taylor}, \citenamefont {Watt},
  \citenamefont {Assender},\ and\ \citenamefont {Smith}}]{2016Smith}%
  \BibitemOpen
  \bibfield  {author} {\bibinfo {author} {\bibfnamefont {A.~W.}\ \bibnamefont
  {Powell}}, \bibinfo {author} {\bibfnamefont {D.~M.}\ \bibnamefont {Coles}},
  \bibinfo {author} {\bibfnamefont {R.~A.}\ \bibnamefont {Taylor}}, \bibinfo
  {author} {\bibfnamefont {A.~A.~R.}\ \bibnamefont {Watt}}, \bibinfo {author}
  {\bibfnamefont {H.~E.}\ \bibnamefont {Assender}},\ and\ \bibinfo {author}
  {\bibfnamefont {J.~M.}\ \bibnamefont {Smith}},\ }\bibfield  {title} {\enquote
  {\bibinfo {title} {Plasmonic gas sensing using nanocube patch antennas},}\
  }\href {https://doi.org/https://doi.org/10.1002/adom.201500602} {\bibfield
  {journal} {\bibinfo  {journal} {Advanced Optical Materials}\ }\textbf
  {\bibinfo {volume} {4}},\ \bibinfo {pages} {634--642} (\bibinfo {year}
  {2016})}\BibitemShut {NoStop}%
\bibitem [{\citenamefont {Xu}\ \emph {et~al.}(2015)\citenamefont {Xu},
  \citenamefont {Yang}, \citenamefont {Jin}, \citenamefont {Chen},
  \citenamefont {Fan}, \citenamefont {Luo},\ and\ \citenamefont
  {Shi}}]{Xu2015}%
  \BibitemOpen
  \bibfield  {author} {\bibinfo {author} {\bibfnamefont {D.}~\bibnamefont
  {Xu}}, \bibinfo {author} {\bibfnamefont {S.}~\bibnamefont {Yang}}, \bibinfo
  {author} {\bibfnamefont {Y.}~\bibnamefont {Jin}}, \bibinfo {author}
  {\bibfnamefont {M.}~\bibnamefont {Chen}}, \bibinfo {author} {\bibfnamefont
  {W.}~\bibnamefont {Fan}}, \bibinfo {author} {\bibfnamefont {B.}~\bibnamefont
  {Luo}},\ and\ \bibinfo {author} {\bibfnamefont {W.}~\bibnamefont {Shi}},\
  }\bibfield  {title} {\enquote {\bibinfo {title} {Ag-decorated atao3 (a = k,
  na) nanocube plasmonic photocatalysts with enhanced photocatalytic
  water-splitting properties},}\ }\href
  {https://doi.org/10.1021/acs.langmuir.5b01294} {\bibfield  {journal}
  {\bibinfo  {journal} {Langmuir}\ }\textbf {\bibinfo {volume} {31}},\ \bibinfo
  {pages} {9694--9699} (\bibinfo {year} {2015})}\BibitemShut {NoStop}%
\bibitem [{\citenamefont {Lu}\ \emph {et~al.}(2020)\citenamefont {Lu},
  \citenamefont {Zhang}, \citenamefont {Zhang},\ and\ \citenamefont
  {Xu}}]{Lu2020}%
  \BibitemOpen
  \bibfield  {author} {\bibinfo {author} {\bibfnamefont {W.}~\bibnamefont
  {Lu}}, \bibinfo {author} {\bibfnamefont {Y.}~\bibnamefont {Zhang}}, \bibinfo
  {author} {\bibfnamefont {J.}~\bibnamefont {Zhang}},\ and\ \bibinfo {author}
  {\bibfnamefont {P.}~\bibnamefont {Xu}},\ }\bibfield  {title} {\enquote
  {\bibinfo {title} {Reduction of gas co2 to co with high selectivity by ag
  nanocube-based membrane cathodes in a photoelectrochemical system},}\ }\href
  {https://doi.org/10.1021/acs.iecr.9b06052} {\bibfield  {journal} {\bibinfo
  {journal} {Industrial {\&} Engineering Chemistry Research}\ }\textbf
  {\bibinfo {volume} {59}},\ \bibinfo {pages} {5536--5545} (\bibinfo {year}
  {2020})}\BibitemShut {NoStop}%
\bibitem [{\citenamefont {Bischak}\ \emph {et~al.}(2017)\citenamefont
  {Bischak}, \citenamefont {Wai}, \citenamefont {Cherqui}, \citenamefont
  {Busche}, \citenamefont {Quillin}, \citenamefont {Hetherington},
  \citenamefont {Wang}, \citenamefont {Aiello}, \citenamefont {Schlom},
  \citenamefont {Aloni}, \citenamefont {Ogletree}, \citenamefont {Masiello},\
  and\ \citenamefont {Ginsberg}}]{Bischak2017}%
  \BibitemOpen
  \bibfield  {author} {\bibinfo {author} {\bibfnamefont {C.~G.}\ \bibnamefont
  {Bischak}}, \bibinfo {author} {\bibfnamefont {R.~B.}\ \bibnamefont {Wai}},
  \bibinfo {author} {\bibfnamefont {C.}~\bibnamefont {Cherqui}}, \bibinfo
  {author} {\bibfnamefont {J.~A.}\ \bibnamefont {Busche}}, \bibinfo {author}
  {\bibfnamefont {S.~C.}\ \bibnamefont {Quillin}}, \bibinfo {author}
  {\bibfnamefont {C.~L.}\ \bibnamefont {Hetherington}}, \bibinfo {author}
  {\bibfnamefont {Z.}~\bibnamefont {Wang}}, \bibinfo {author} {\bibfnamefont
  {C.~D.}\ \bibnamefont {Aiello}}, \bibinfo {author} {\bibfnamefont {D.~G.}\
  \bibnamefont {Schlom}}, \bibinfo {author} {\bibfnamefont {S.}~\bibnamefont
  {Aloni}}, \bibinfo {author} {\bibfnamefont {D.~F.}\ \bibnamefont {Ogletree}},
  \bibinfo {author} {\bibfnamefont {D.~J.}\ \bibnamefont {Masiello}},\ and\
  \bibinfo {author} {\bibfnamefont {N.~S.}\ \bibnamefont {Ginsberg}},\
  }\bibfield  {title} {\enquote {\bibinfo {title} {Noninvasive
  cathodoluminescence-activated nanoimaging of dynamic processes in liquids},}\
  }\href {https://doi.org/10.1021/acsnano.7b06081} {\bibfield  {journal}
  {\bibinfo  {journal} {ACS Nano}\ }\textbf {\bibinfo {volume} {11}},\ \bibinfo
  {pages} {10583--10590} (\bibinfo {year} {2017})}\BibitemShut {NoStop}%
\bibitem [{\citenamefont {Hoang}\ and\ \citenamefont
  {Mikkelsen}(2016)}]{ElecTuning2016}%
  \BibitemOpen
  \bibfield  {author} {\bibinfo {author} {\bibfnamefont {T.~B.}\ \bibnamefont
  {Hoang}}\ and\ \bibinfo {author} {\bibfnamefont {M.~H.}\ \bibnamefont
  {Mikkelsen}},\ }\bibfield  {title} {\enquote {\bibinfo {title} {Broad
  electrical tuning of plasmonic nanoantennas at visible frequencies},}\ }\href
  {https://doi.org/10.1063/1.4948588} {\bibfield  {journal} {\bibinfo
  {journal} {Applied Physics Letters}\ }\textbf {\bibinfo {volume} {108}},\
  \bibinfo {pages} {183107} (\bibinfo {year} {2016})}\BibitemShut {NoStop}%
\bibitem [{\citenamefont {Hachtel}\ \emph {et~al.}(2019)\citenamefont
  {Hachtel}, \citenamefont {Cho}, \citenamefont {Davidson}, \citenamefont
  {Feldman}, \citenamefont {Chisholm}, \citenamefont {Haglund}, \citenamefont
  {Idrobo}, \citenamefont {Pantelides},\ and\ \citenamefont
  {Lawrie}}]{Hachtel2019}%
  \BibitemOpen
  \bibfield  {author} {\bibinfo {author} {\bibfnamefont {J.~A.}\ \bibnamefont
  {Hachtel}}, \bibinfo {author} {\bibfnamefont {S.-Y.}\ \bibnamefont {Cho}},
  \bibinfo {author} {\bibfnamefont {R.~B.}\ \bibnamefont {Davidson}}, \bibinfo
  {author} {\bibfnamefont {M.~A.}\ \bibnamefont {Feldman}}, \bibinfo {author}
  {\bibfnamefont {M.~F.}\ \bibnamefont {Chisholm}}, \bibinfo {author}
  {\bibfnamefont {R.~F.}\ \bibnamefont {Haglund}}, \bibinfo {author}
  {\bibfnamefont {J.~C.}\ \bibnamefont {Idrobo}}, \bibinfo {author}
  {\bibfnamefont {S.~T.}\ \bibnamefont {Pantelides}},\ and\ \bibinfo {author}
  {\bibfnamefont {B.~J.}\ \bibnamefont {Lawrie}},\ }\bibfield  {title}
  {\enquote {\bibinfo {title} {Spatially and spectrally resolved orbital
  angular momentum interactions in plasmonic vortex generators},}\ }\href
  {https://doi.org/10.1038/s41377-019-0136-z} {\bibfield  {journal} {\bibinfo
  {journal} {Light: Science {\&} Applications}\ }\textbf {\bibinfo {volume}
  {8}},\ \bibinfo {pages} {33} (\bibinfo {year} {2019})}\BibitemShut {NoStop}%
\bibitem [{\citenamefont {Pakeltis}\ \emph {et~al.}(2021)\citenamefont
  {Pakeltis}, \citenamefont {Rotunno}, \citenamefont {Khorassani},
  \citenamefont {Garfinkel}, \citenamefont {Collette}, \citenamefont {West},
  \citenamefont {Retterer}, \citenamefont {Idrobo}, \citenamefont {Masiello},\
  and\ \citenamefont {Rack}}]{Pakeltis:21}%
  \BibitemOpen
  \bibfield  {author} {\bibinfo {author} {\bibfnamefont {G.}~\bibnamefont
  {Pakeltis}}, \bibinfo {author} {\bibfnamefont {E.}~\bibnamefont {Rotunno}},
  \bibinfo {author} {\bibfnamefont {S.}~\bibnamefont {Khorassani}}, \bibinfo
  {author} {\bibfnamefont {D.~A.}\ \bibnamefont {Garfinkel}}, \bibinfo {author}
  {\bibfnamefont {R.}~\bibnamefont {Collette}}, \bibinfo {author}
  {\bibfnamefont {C.~A.}\ \bibnamefont {West}}, \bibinfo {author}
  {\bibfnamefont {S.~T.}\ \bibnamefont {Retterer}}, \bibinfo {author}
  {\bibfnamefont {J.~C.}\ \bibnamefont {Idrobo}}, \bibinfo {author}
  {\bibfnamefont {D.~J.}\ \bibnamefont {Masiello}},\ and\ \bibinfo {author}
  {\bibfnamefont {P.~D.}\ \bibnamefont {Rack}},\ }\bibfield  {title} {\enquote
  {\bibinfo {title} {High spatial and energy resolution electron energy loss
  spectroscopy of the magnetic and electric excitations in plasmonic nanorod
  oligomers},}\ }\href {https://doi.org/10.1364/OE.416046} {\bibfield
  {journal} {\bibinfo  {journal} {Opt. Express}\ }\textbf {\bibinfo {volume}
  {29}},\ \bibinfo {pages} {4661--4671} (\bibinfo {year} {2021})}\BibitemShut
  {NoStop}%
\bibitem [{\citenamefont {Hachtel}\ \emph {et~al.}(2018)\citenamefont
  {Hachtel}, \citenamefont {Davidson}, \citenamefont {Kovalik}, \citenamefont
  {Retterer}, \citenamefont {Lupini}, \citenamefont {Haglund}, \citenamefont
  {Lawrie},\ and\ \citenamefont {Pantelides}}]{hachtel2018polarization}%
  \BibitemOpen
  \bibfield  {author} {\bibinfo {author} {\bibfnamefont {J.~A.}\ \bibnamefont
  {Hachtel}}, \bibinfo {author} {\bibfnamefont {R.~B.}\ \bibnamefont
  {Davidson}}, \bibinfo {author} {\bibfnamefont {E.~R.}\ \bibnamefont
  {Kovalik}}, \bibinfo {author} {\bibfnamefont {S.~T.}\ \bibnamefont
  {Retterer}}, \bibinfo {author} {\bibfnamefont {A.~R.}\ \bibnamefont
  {Lupini}}, \bibinfo {author} {\bibfnamefont {R.~F.}\ \bibnamefont {Haglund}},
  \bibinfo {author} {\bibfnamefont {B.~J.}\ \bibnamefont {Lawrie}},\ and\
  \bibinfo {author} {\bibfnamefont {S.~T.}\ \bibnamefont {Pantelides}},\
  }\bibfield  {title} {\enquote {\bibinfo {title} {Polarization-and
  wavelength-resolved near-field imaging of complex plasmonic modes in
  archimedean nanospirals},}\ }\href@noop {} {\bibfield  {journal} {\bibinfo
  {journal} {Optics letters}\ }\textbf {\bibinfo {volume} {43}},\ \bibinfo
  {pages} {927--930} (\bibinfo {year} {2018})}\BibitemShut {NoStop}%
\bibitem [{\citenamefont {Feldman}\ \emph {et~al.}(2018)\citenamefont
  {Feldman}, \citenamefont {Dumitrescu}, \citenamefont {Bridges}, \citenamefont
  {Chisholm}, \citenamefont {Davidson}, \citenamefont {Evans}, \citenamefont
  {Hachtel}, \citenamefont {Hu}, \citenamefont {Pooser}, \citenamefont
  {Haglund},\ and\ \citenamefont {Lawrie}}]{feldman2018colossal}%
  \BibitemOpen
  \bibfield  {author} {\bibinfo {author} {\bibfnamefont {M.~A.}\ \bibnamefont
  {Feldman}}, \bibinfo {author} {\bibfnamefont {E.~F.}\ \bibnamefont
  {Dumitrescu}}, \bibinfo {author} {\bibfnamefont {D.}~\bibnamefont {Bridges}},
  \bibinfo {author} {\bibfnamefont {M.~F.}\ \bibnamefont {Chisholm}}, \bibinfo
  {author} {\bibfnamefont {R.~B.}\ \bibnamefont {Davidson}}, \bibinfo {author}
  {\bibfnamefont {P.~G.}\ \bibnamefont {Evans}}, \bibinfo {author}
  {\bibfnamefont {J.~A.}\ \bibnamefont {Hachtel}}, \bibinfo {author}
  {\bibfnamefont {A.}~\bibnamefont {Hu}}, \bibinfo {author} {\bibfnamefont
  {R.~C.}\ \bibnamefont {Pooser}}, \bibinfo {author} {\bibfnamefont {R.~F.}\
  \bibnamefont {Haglund}},\ and\ \bibinfo {author} {\bibfnamefont {B.~J.}\
  \bibnamefont {Lawrie}},\ }\bibfield  {title} {\enquote {\bibinfo {title}
  {Colossal photon bunching in quasiparticle-mediated nanodiamond
  cathodoluminescence},}\ }\href {https://doi.org/10.1103/PhysRevB.97.081404}
  {\bibfield  {journal} {\bibinfo  {journal} {Phys. Rev. B}\ }\textbf {\bibinfo
  {volume} {97}},\ \bibinfo {pages} {081404} (\bibinfo {year}
  {2018})}\BibitemShut {NoStop}%
\bibitem [{\citenamefont {Goris}\ \emph {et~al.}(2014)\citenamefont {Goris},
  \citenamefont {Guzzinati}, \citenamefont {Fern{\'a}ndez-L{\'o}pez},
  \citenamefont {P{\'e}rez-Juste}, \citenamefont {Liz-Marz{\'a}n},
  \citenamefont {Tr{\"u}gler}, \citenamefont {Hohenester}, \citenamefont
  {Verbeeck}, \citenamefont {Bals},\ and\ \citenamefont
  {Van~Tendeloo}}]{Goris2014}%
  \BibitemOpen
  \bibfield  {author} {\bibinfo {author} {\bibfnamefont {B.}~\bibnamefont
  {Goris}}, \bibinfo {author} {\bibfnamefont {G.}~\bibnamefont {Guzzinati}},
  \bibinfo {author} {\bibfnamefont {C.}~\bibnamefont
  {Fern{\'a}ndez-L{\'o}pez}}, \bibinfo {author} {\bibfnamefont
  {J.}~\bibnamefont {P{\'e}rez-Juste}}, \bibinfo {author} {\bibfnamefont
  {L.~M.}\ \bibnamefont {Liz-Marz{\'a}n}}, \bibinfo {author} {\bibfnamefont
  {A.}~\bibnamefont {Tr{\"u}gler}}, \bibinfo {author} {\bibfnamefont
  {U.}~\bibnamefont {Hohenester}}, \bibinfo {author} {\bibfnamefont
  {J.}~\bibnamefont {Verbeeck}}, \bibinfo {author} {\bibfnamefont
  {S.}~\bibnamefont {Bals}},\ and\ \bibinfo {author} {\bibfnamefont
  {G.}~\bibnamefont {Van~Tendeloo}},\ }\bibfield  {title} {\enquote {\bibinfo
  {title} {Plasmon mapping in au@ag nanocube assemblies},}\ }\href
  {https://doi.org/10.1021/jp502584t} {\bibfield  {journal} {\bibinfo
  {journal} {The Journal of Physical Chemistry C}\ }\textbf {\bibinfo {volume}
  {118}},\ \bibinfo {pages} {15356--15362} (\bibinfo {year}
  {2014})}\BibitemShut {NoStop}%
\bibitem [{\citenamefont {Kone{\v{c}}n{\'a}}\ \emph {et~al.}(2021)\citenamefont
  {Kone{\v{c}}n{\'a}}, \citenamefont {Li}, \citenamefont {Edgar}, \citenamefont
  {de~Abajo},\ and\ \citenamefont {Hachtel}}]{konevcna2021revealing}%
  \BibitemOpen
  \bibfield  {author} {\bibinfo {author} {\bibfnamefont {A.}~\bibnamefont
  {Kone{\v{c}}n{\'a}}}, \bibinfo {author} {\bibfnamefont {J.}~\bibnamefont
  {Li}}, \bibinfo {author} {\bibfnamefont {J.~H.}\ \bibnamefont {Edgar}},
  \bibinfo {author} {\bibfnamefont {F.}~\bibnamefont {de~Abajo}},\ and\
  \bibinfo {author} {\bibfnamefont {J.~A.}\ \bibnamefont {Hachtel}},\
  }\bibfield  {title} {\enquote {\bibinfo {title} {Revealing nanoscale
  confinement effects on hyperbolic phonon polaritons with an electron beam},}\
  }\href@noop {} {\bibfield  {journal} {\bibinfo  {journal} {arXiv preprint
  arXiv:2104.01997}\ } (\bibinfo {year} {2021})}\BibitemShut {NoStop}%
\bibitem [{\citenamefont {Liu}\ \emph {et~al.}(2019)\citenamefont {Liu},
  \citenamefont {Guo}, \citenamefont {Ramoji}, \citenamefont {Bocklitz},
  \citenamefont {Rosch}, \citenamefont {Popp},\ and\ \citenamefont
  {Yu}}]{liu2019spatiotemporal}%
  \BibitemOpen
  \bibfield  {author} {\bibinfo {author} {\bibfnamefont {X.-Y.}\ \bibnamefont
  {Liu}}, \bibinfo {author} {\bibfnamefont {S.}~\bibnamefont {Guo}}, \bibinfo
  {author} {\bibfnamefont {A.}~\bibnamefont {Ramoji}}, \bibinfo {author}
  {\bibfnamefont {T.}~\bibnamefont {Bocklitz}}, \bibinfo {author}
  {\bibfnamefont {P.}~\bibnamefont {Rosch}}, \bibinfo {author} {\bibfnamefont
  {J.}~\bibnamefont {Popp}},\ and\ \bibinfo {author} {\bibfnamefont {H.-Q.}\
  \bibnamefont {Yu}},\ }\bibfield  {title} {\enquote {\bibinfo {title}
  {Spatiotemporal organization of biofilm matrix revealed by confocal raman
  mapping integrated with non-negative matrix factorization analysis},}\
  }\href@noop {} {\bibfield  {journal} {\bibinfo  {journal} {Analytical
  chemistry}\ }\textbf {\bibinfo {volume} {92}},\ \bibinfo {pages} {707--715}
  (\bibinfo {year} {2019})}\BibitemShut {NoStop}%
\bibitem [{\citenamefont {Sannomiya}\ \emph {et~al.}(2020)\citenamefont
  {Sannomiya}, \citenamefont {Kone{\v{c}}n{\'a}}, \citenamefont {Matsukata},
  \citenamefont {Thollar}, \citenamefont {Okamoto}, \citenamefont
  {Garc{\'i}a~de Abajo},\ and\ \citenamefont {Yamamoto}}]{Sannomiya2020}%
  \BibitemOpen
  \bibfield  {author} {\bibinfo {author} {\bibfnamefont {T.}~\bibnamefont
  {Sannomiya}}, \bibinfo {author} {\bibfnamefont {A.}~\bibnamefont
  {Kone{\v{c}}n{\'a}}}, \bibinfo {author} {\bibfnamefont {T.}~\bibnamefont
  {Matsukata}}, \bibinfo {author} {\bibfnamefont {Z.}~\bibnamefont {Thollar}},
  \bibinfo {author} {\bibfnamefont {T.}~\bibnamefont {Okamoto}}, \bibinfo
  {author} {\bibfnamefont {F.~J.}\ \bibnamefont {Garc{\'i}a~de Abajo}},\ and\
  \bibinfo {author} {\bibfnamefont {N.}~\bibnamefont {Yamamoto}},\ }\bibfield
  {title} {\enquote {\bibinfo {title} {Cathodoluminescence phase extraction of
  the coupling between nanoparticles and surface plasmon polaritons},}\ }\href
  {https://doi.org/10.1021/acs.nanolett.9b04335} {\bibfield  {journal}
  {\bibinfo  {journal} {Nano Letters}\ }\textbf {\bibinfo {volume} {20}},\
  \bibinfo {pages} {592--598} (\bibinfo {year} {2020})}\BibitemShut {NoStop}%
\bibitem [{\citenamefont {Kuttge}\ \emph {et~al.}(2009)\citenamefont {Kuttge},
  \citenamefont {Vesseur}, \citenamefont {Koenderink}, \citenamefont {Lezec},
  \citenamefont {Atwater}, \citenamefont {de~Abajo},\ and\ \citenamefont
  {Polman}}]{kuttge2009local}%
  \BibitemOpen
  \bibfield  {author} {\bibinfo {author} {\bibfnamefont {M.}~\bibnamefont
  {Kuttge}}, \bibinfo {author} {\bibfnamefont {E.~J.~R.}\ \bibnamefont
  {Vesseur}}, \bibinfo {author} {\bibfnamefont {A.}~\bibnamefont {Koenderink}},
  \bibinfo {author} {\bibfnamefont {H.}~\bibnamefont {Lezec}}, \bibinfo
  {author} {\bibfnamefont {H.}~\bibnamefont {Atwater}}, \bibinfo {author}
  {\bibfnamefont {F.~G.}\ \bibnamefont {de~Abajo}},\ and\ \bibinfo {author}
  {\bibfnamefont {A.}~\bibnamefont {Polman}},\ }\bibfield  {title} {\enquote
  {\bibinfo {title} {Local density of states, spectrum, and far-field
  interference of surface plasmon polaritons probed by cathodoluminescence},}\
  }\href@noop {} {\bibfield  {journal} {\bibinfo  {journal} {Physical Review
  B}\ }\textbf {\bibinfo {volume} {79}},\ \bibinfo {pages} {113405} (\bibinfo
  {year} {2009})}\BibitemShut {NoStop}%
\bibitem [{\citenamefont {Dumitrescu}\ and\ \citenamefont
  {Lawrie}(2017)}]{dumitrescu2017antibunching}%
  \BibitemOpen
  \bibfield  {author} {\bibinfo {author} {\bibfnamefont {E.}~\bibnamefont
  {Dumitrescu}}\ and\ \bibinfo {author} {\bibfnamefont {B.}~\bibnamefont
  {Lawrie}},\ }\bibfield  {title} {\enquote {\bibinfo {title} {Antibunching
  dynamics of plasmonically mediated entanglement generation},}\ }\href@noop {}
  {\bibfield  {journal} {\bibinfo  {journal} {Physical Review A}\ }\textbf
  {\bibinfo {volume} {96}},\ \bibinfo {pages} {053826} (\bibinfo {year}
  {2017})}\BibitemShut {NoStop}%
\bibitem [{\citenamefont {Li}, \citenamefont {Nemilentsau},\ and\ \citenamefont
  {Argyropoulos}(2019)}]{li2019resonance}%
  \BibitemOpen
  \bibfield  {author} {\bibinfo {author} {\bibfnamefont {Y.}~\bibnamefont
  {Li}}, \bibinfo {author} {\bibfnamefont {A.}~\bibnamefont {Nemilentsau}},\
  and\ \bibinfo {author} {\bibfnamefont {C.}~\bibnamefont {Argyropoulos}},\
  }\bibfield  {title} {\enquote {\bibinfo {title} {Resonance energy transfer
  and quantum entanglement mediated by epsilon-near-zero and other plasmonic
  waveguide systems},}\ }\href@noop {} {\bibfield  {journal} {\bibinfo
  {journal} {Nanoscale}\ }\textbf {\bibinfo {volume} {11}},\ \bibinfo {pages}
  {14635--14647} (\bibinfo {year} {2019})}\BibitemShut {NoStop}%
\bibitem [{\citenamefont {Jiang}\ \emph {et~al.}(2020)\citenamefont {Jiang},
  \citenamefont {Zheng}, \citenamefont {Li}, \citenamefont {Shan},
  \citenamefont {Chi}, \citenamefont {Liu}, \citenamefont {Huang},
  \citenamefont {Dang}, \citenamefont {Lin},\ and\ \citenamefont
  {Fang}}]{jiang2020tailoring}%
  \BibitemOpen
  \bibfield  {author} {\bibinfo {author} {\bibfnamefont {M.}~\bibnamefont
  {Jiang}}, \bibinfo {author} {\bibfnamefont {L.}~\bibnamefont {Zheng}},
  \bibinfo {author} {\bibfnamefont {Y.}~\bibnamefont {Li}}, \bibinfo {author}
  {\bibfnamefont {H.}~\bibnamefont {Shan}}, \bibinfo {author} {\bibfnamefont
  {C.}~\bibnamefont {Chi}}, \bibinfo {author} {\bibfnamefont {Z.}~\bibnamefont
  {Liu}}, \bibinfo {author} {\bibfnamefont {Y.}~\bibnamefont {Huang}}, \bibinfo
  {author} {\bibfnamefont {Z.}~\bibnamefont {Dang}}, \bibinfo {author}
  {\bibfnamefont {F.}~\bibnamefont {Lin}},\ and\ \bibinfo {author}
  {\bibfnamefont {Z.}~\bibnamefont {Fang}},\ }\bibfield  {title} {\enquote
  {\bibinfo {title} {Tailoring zno spontaneous emission with plasmonic
  radiative local density of states using cathodoluminescence microscopy},}\
  }\href@noop {} {\bibfield  {journal} {\bibinfo  {journal} {The Journal of
  Physical Chemistry C}\ }\textbf {\bibinfo {volume} {124}},\ \bibinfo {pages}
  {13886--13893} (\bibinfo {year} {2020})}\BibitemShut {NoStop}%
\bibitem [{\citenamefont {Yanagimoto}\ \emph {et~al.}(2020)\citenamefont
  {Yanagimoto}, \citenamefont {Yamamoto}, \citenamefont {Sannomiya},\ and\
  \citenamefont {Akiba}}]{yanagimoto2020purcell}%
  \BibitemOpen
  \bibfield  {author} {\bibinfo {author} {\bibfnamefont {S.}~\bibnamefont
  {Yanagimoto}}, \bibinfo {author} {\bibfnamefont {N.}~\bibnamefont
  {Yamamoto}}, \bibinfo {author} {\bibfnamefont {T.}~\bibnamefont
  {Sannomiya}},\ and\ \bibinfo {author} {\bibfnamefont {K.}~\bibnamefont
  {Akiba}},\ }\bibfield  {title} {\enquote {\bibinfo {title} {Purcell effect of
  nitrogen-vacancy centers in nanodiamond coupled to propagating and localized
  surface plasmons revealed by photon-correlation cathodoluminescence},}\
  }\href@noop {} {\bibfield  {journal} {\bibinfo  {journal} {arXiv preprint
  arXiv:2012.11224}\ } (\bibinfo {year} {2020})}\BibitemShut {NoStop}%
\bibitem [{\citenamefont {Hoang}\ \emph {et~al.}(2015)\citenamefont {Hoang},
  \citenamefont {Akselrod}, \citenamefont {Argyropoulos}, \citenamefont
  {Huang}, \citenamefont {Smith},\ and\ \citenamefont
  {Mikkelsen}}]{Hoang2015UF}%
  \BibitemOpen
  \bibfield  {author} {\bibinfo {author} {\bibfnamefont {T.~B.}\ \bibnamefont
  {Hoang}}, \bibinfo {author} {\bibfnamefont {G.~M.}\ \bibnamefont {Akselrod}},
  \bibinfo {author} {\bibfnamefont {C.}~\bibnamefont {Argyropoulos}}, \bibinfo
  {author} {\bibfnamefont {J.}~\bibnamefont {Huang}}, \bibinfo {author}
  {\bibfnamefont {D.~R.}\ \bibnamefont {Smith}},\ and\ \bibinfo {author}
  {\bibfnamefont {M.~H.}\ \bibnamefont {Mikkelsen}},\ }\bibfield  {title}
  {\enquote {\bibinfo {title} {Ultrafast spontaneous emission source using
  plasmonic nanoantennas},}\ }\href {https://doi.org/10.1038/ncomms8788}
  {\bibfield  {journal} {\bibinfo  {journal} {Nature Communications}\ }\textbf
  {\bibinfo {volume} {6}},\ \bibinfo {pages} {7788} (\bibinfo {year}
  {2015})}\BibitemShut {NoStop}%
\bibitem [{\citenamefont {Winkler}\ \emph {et~al.}(2017)\citenamefont
  {Winkler}, \citenamefont {Schmidt}, \citenamefont {Haselmann}, \citenamefont
  {Fowlkes}, \citenamefont {Lewis}, \citenamefont {Kothleitner}, \citenamefont
  {Rack},\ and\ \citenamefont {Plank}}]{winkler2017direct}%
  \BibitemOpen
  \bibfield  {author} {\bibinfo {author} {\bibfnamefont {R.}~\bibnamefont
  {Winkler}}, \bibinfo {author} {\bibfnamefont {F.-P.}\ \bibnamefont
  {Schmidt}}, \bibinfo {author} {\bibfnamefont {U.}~\bibnamefont {Haselmann}},
  \bibinfo {author} {\bibfnamefont {J.~D.}\ \bibnamefont {Fowlkes}}, \bibinfo
  {author} {\bibfnamefont {B.~B.}\ \bibnamefont {Lewis}}, \bibinfo {author}
  {\bibfnamefont {G.}~\bibnamefont {Kothleitner}}, \bibinfo {author}
  {\bibfnamefont {P.~D.}\ \bibnamefont {Rack}},\ and\ \bibinfo {author}
  {\bibfnamefont {H.}~\bibnamefont {Plank}},\ }\bibfield  {title} {\enquote
  {\bibinfo {title} {Direct-write 3d nanoprinting of plasmonic structures},}\
  }\href@noop {} {\bibfield  {journal} {\bibinfo  {journal} {ACS applied
  materials \& interfaces}\ }\textbf {\bibinfo {volume} {9}},\ \bibinfo {pages}
  {8233--8240} (\bibinfo {year} {2017})}\BibitemShut {NoStop}%
\bibitem [{\citenamefont {Iyer}\ \emph {et~al.}(2021)\citenamefont {Iyer},
  \citenamefont {Retterer}, \citenamefont {Fowlkes}, \citenamefont {Jesse},
  \citenamefont {Puretzky}, \citenamefont {Hachtel}, \citenamefont {Rack},\
  and\ \citenamefont {Lawrie}}]{iyer2021situ}%
  \BibitemOpen
  \bibfield  {author} {\bibinfo {author} {\bibfnamefont {V.}~\bibnamefont
  {Iyer}}, \bibinfo {author} {\bibfnamefont {S.~T.}\ \bibnamefont {Retterer}},
  \bibinfo {author} {\bibfnamefont {J.}~\bibnamefont {Fowlkes}}, \bibinfo
  {author} {\bibfnamefont {S.}~\bibnamefont {Jesse}}, \bibinfo {author}
  {\bibfnamefont {A.~A.}\ \bibnamefont {Puretzky}}, \bibinfo {author}
  {\bibfnamefont {J.~A.}\ \bibnamefont {Hachtel}}, \bibinfo {author}
  {\bibfnamefont {P.~D.}\ \bibnamefont {Rack}},\ and\ \bibinfo {author}
  {\bibfnamefont {B.~J.}\ \bibnamefont {Lawrie}},\ }\bibfield  {title}
  {\enquote {\bibinfo {title} {In situ electron-beam processing and
  cathodoluminescence microscopy for quantum nanophotonics},}\ }in\ \href@noop
  {} {\emph {\bibinfo {booktitle} {Synthesis and Photonics of Nanoscale
  Materials XVIII}}},\ Vol.\ \bibinfo {volume} {11675}\ (\bibinfo
  {organization} {International Society for Optics and Photonics},\ \bibinfo
  {year} {2021})\ p.\ \bibinfo {pages} {1167504}\BibitemShut {NoStop}%
\bibitem [{\citenamefont {Sutter}\ \emph
  {et~al.}(2021{\natexlab{a}})\citenamefont {Sutter}, \citenamefont
  {Khorashad}, \citenamefont {Argyropoulos},\ and\ \citenamefont
  {Sutter}}]{Sutter2021}%
  \BibitemOpen
  \bibfield  {author} {\bibinfo {author} {\bibfnamefont {P.}~\bibnamefont
  {Sutter}}, \bibinfo {author} {\bibfnamefont {L.~K.}\ \bibnamefont
  {Khorashad}}, \bibinfo {author} {\bibfnamefont {C.}~\bibnamefont
  {Argyropoulos}},\ and\ \bibinfo {author} {\bibfnamefont {E.}~\bibnamefont
  {Sutter}},\ }\bibfield  {title} {\enquote {\bibinfo {title}
  {Cathodoluminescence of ultrathin twisted ge1–xsnxs van der waals
  nanoribbon waveguides},}\ }\href
  {https://doi.org/https://doi.org/10.1002/adma.202006649} {\bibfield
  {journal} {\bibinfo  {journal} {Advanced Materials}\ }\textbf {\bibinfo
  {volume} {33}},\ \bibinfo {pages} {2006649} (\bibinfo {year}
  {2021}{\natexlab{a}})}\BibitemShut {NoStop}%
\bibitem [{\citenamefont {Sutter}\ \emph
  {et~al.}(2021{\natexlab{b}})\citenamefont {Sutter}, \citenamefont {French},
  \citenamefont {Khosravi~Khorashad}, \citenamefont {Argyropoulos},\ and\
  \citenamefont {Sutter}}]{Sutter2021Opto}%
  \BibitemOpen
  \bibfield  {author} {\bibinfo {author} {\bibfnamefont {P.}~\bibnamefont
  {Sutter}}, \bibinfo {author} {\bibfnamefont {J.~S.}\ \bibnamefont {French}},
  \bibinfo {author} {\bibfnamefont {L.}~\bibnamefont {Khosravi~Khorashad}},
  \bibinfo {author} {\bibfnamefont {C.}~\bibnamefont {Argyropoulos}},\ and\
  \bibinfo {author} {\bibfnamefont {E.}~\bibnamefont {Sutter}},\ }\bibfield
  {title} {\enquote {\bibinfo {title} {Optoelectronics and nanophotonics of
  vapor--liquid--solid grown gase van der waals nanoribbons},}\ }\href
  {https://doi.org/10.1021/acs.nanolett.1c00891} {\bibfield  {journal}
  {\bibinfo  {journal} {Nano Letters}\ }\textbf {\bibinfo {volume} {21}},\
  \bibinfo {pages} {4335--4342} (\bibinfo {year}
  {2021}{\natexlab{b}})}\BibitemShut {NoStop}%
\bibitem [{\citenamefont {Stelling}\ \emph {et~al.}(2017)\citenamefont
  {Stelling}, \citenamefont {Singh}, \citenamefont {Karg}, \citenamefont
  {K{\"o}nig}, \citenamefont {Thelakkat},\ and\ \citenamefont
  {Retsch}}]{Stelling2017}%
  \BibitemOpen
  \bibfield  {author} {\bibinfo {author} {\bibfnamefont {C.}~\bibnamefont
  {Stelling}}, \bibinfo {author} {\bibfnamefont {C.~R.}\ \bibnamefont {Singh}},
  \bibinfo {author} {\bibfnamefont {M.}~\bibnamefont {Karg}}, \bibinfo {author}
  {\bibfnamefont {T.~A.~F.}\ \bibnamefont {K{\"o}nig}}, \bibinfo {author}
  {\bibfnamefont {M.}~\bibnamefont {Thelakkat}},\ and\ \bibinfo {author}
  {\bibfnamefont {M.}~\bibnamefont {Retsch}},\ }\bibfield  {title} {\enquote
  {\bibinfo {title} {Plasmonic nanomeshes: their ambivalent role as transparent
  electrodes in organic solar cells},}\ }\href
  {https://doi.org/10.1038/srep42530} {\bibfield  {journal} {\bibinfo
  {journal} {Scientific Reports}\ }\textbf {\bibinfo {volume} {7}},\ \bibinfo
  {pages} {42530} (\bibinfo {year} {2017})}\BibitemShut {NoStop}%
\bibitem [{\citenamefont {Johnson}\ and\ \citenamefont
  {Christy}(1972)}]{PhysRevB.6.4370}%
  \BibitemOpen
  \bibfield  {author} {\bibinfo {author} {\bibfnamefont {P.~B.}\ \bibnamefont
  {Johnson}}\ and\ \bibinfo {author} {\bibfnamefont {R.~W.}\ \bibnamefont
  {Christy}},\ }\bibfield  {title} {\enquote {\bibinfo {title} {Optical
  constants of the noble metals},}\ }\href
  {https://doi.org/10.1103/PhysRevB.6.4370} {\bibfield  {journal} {\bibinfo
  {journal} {Phys. Rev. B}\ }\textbf {\bibinfo {volume} {6}},\ \bibinfo {pages}
  {4370--4379} (\bibinfo {year} {1972})}\BibitemShut {NoStop}%
\end{thebibliography}

%

\end{document}